\documentclass[%
 reprint,
superscriptaddress,
 amsmath,amssymb,
 aps,
pre,
]{revtex4-1}

\usepackage{graphicx}
\usepackage{dcolumn}
\usepackage{bm}

\begin{document}

\preprint{APS/123-QED}

\title{Driving-assisted open quantum transport in qubit networks}

\author{Donny \surname{Dwiputra}}
 \email{donny.dwiputra@s.itb.ac.id}
 \affiliation{Theoretical Physics Laboratory, Faculty of Mathematics and Natural Sciences, Institut Teknologi Bandung, Jl. Ganesha 10, Bandung 40132, Indonesia}
\author{Jusak S. \surname{Kosasih}}%
 \affiliation{Theoretical Physics Laboratory, Faculty of Mathematics and Natural Sciences, Institut Teknologi Bandung, Jl. Ganesha 10, Bandung 40132, Indonesia}
 \affiliation{Indonesian Center for Theoretical and Mathematical Physics (ICTMP), Indonesia}
\author{Albertus \surname{Sulaiman}}
 \affiliation{Indonesian Center for Theoretical and Mathematical Physics (ICTMP), Indonesia}
  \affiliation{Badan Pengkajian dan Penerapan Teknologi, BPPT Bld. II (19 th floor), \\ Jl. M.H. Thamrin 8, Jakarta 10340, Indonesia}
\author{Freddy P. \surname{Zen}}
\email{fpzen@fi.itb.ac.id}
 \affiliation{Theoretical Physics Laboratory, Faculty of Mathematics and Natural Sciences, Institut Teknologi Bandung, Jl. Ganesha 10, Bandung 40132, Indonesia}
 \affiliation{Indonesian Center for Theoretical and Mathematical Physics (ICTMP), Bandung 40132, Indonesia}

\date{\today}

\begin{abstract}
We determine the characteristic of dissipative quantum transport in a coupled qubit network in the presence of on-site and off-diagonal external driving. The work is a generalization of the dephasing-assisted quantum transport where noise is beneficial to the transport efficiency. Using Floquet-Magnus expansion extended to Markovian open systems, we analytically derive transport efficiency and compare it to exact numerical results. We find that periodic driving may increase the efficiency at frequencies near the coupling rate. On the other hand, at some other frequencies the transport may be suppressed. We then propose the enhancement mechanism as the ramification of interplay between driving frequency, dissipative, and trapping rates.
\end{abstract}

\maketitle


\section{\label{sec:intro}INTRODUCTION}
The properties of quantum systems can be engineered in myriad ways through the application of coherent external driving. This type of quantum engineering is based on the Floquet theorem and in the past few years it has gained much interest in both experiment and theory \cite{liu2013floquet,haddadfarshi2015completely,eisert2015quantum,sieberer2016keldysh,restrepo2016driven,shu2018observation,dai2016floquet}. Particularly, in the field of quantum transport, the efficient transport of optical excitation through a network of coupled many-body quantum systems has been extensively studied in both natural and artificial systems \cite{behzadi2015dephasing,barmettler2008quantum,wall2015quantum}. The role of periodic driving is to modify the many-body system of interest so that it presents certain desired properties. For example, near-resonant periodic driving of many-body quantum systems can be applied to demonstrate full control of the Floquet state population \cite{desbuquois2017controlling}. Given the potential of such Floquet engineering, it has now become one of the tools to realize quantum simulators, which paves the way to understand the complex and inaccessible many-body phenomena \cite{kyriienko2018floquet}.

In the past decade, theoretical approaches have demonstrated the potentially beneficial role of noise in quantum transport \cite{plenio2008dephasing,rebentrost2009environment,chin2012coherence,kassal2012environment}, usually in the spirit of delocalized excitons in natural light-harvesting complexes \cite{jang2018delocalized}. That is, the existence of noise in the warm and wet environment of photosynthetic complexes can increase the transport efficiency instead of suppressing it, a phenomenon called environment-assisted quantum transport (ENAQT). It is explained by a model which usually consists of a network of coupled two-state systems (qubits) in a thermal bath. The mechanism of ENAQT is initially understood as a result of the destruction of Anderson localization \cite{mohseni2008environment,plenio2008dephasing,rebentrost2009environment} in a disordered system (having different energies) by dephasing noise. One may think that ENAQT ceases to exist in an ordered system, but it is shown in Ref. \cite{kassal2012environment} that the ENAQT is impossible only for end-to-end transport in an ordered linear chain. Thus the aforementioned mechanism is not the whole story. Instead, it should include the interference effects due to the interplay between dephasing, dissipation, and trapping rates. Furthermore, in some specific scenarios ENAQT can also be viewed as a momentum rejuvenation to counter the broad momentum distribution induced by classical noise \cite{li2015momentum}.

However, the existence of the long-lived electronic quantum coherence, which was initially thought to be responsible in the delocalized excitonic transport \cite{engel2007evidence,panitchayangkoon2010long}, is disputed in a recent experiment \cite{duan2017nature}. Nevertheless, ENAQT is evident in the system and has been studied in recent experiments of environment engineering in the spirit of quantum simulations \cite{sowa2017environment,biggerstaff2016enhancing,trautmann2018trapped,schempp2015correlated,maier2019environment}.

In this paper, we extend the qubit network model of ENAQT to contain external driving in the spirit of Floquet engineering. It should be noted that our model is not intended to extend the understanding of biological energy transport \textit{in vivo}. Instead, our aim is to illustrate phenomena in driven-dissipative quantum simulations, since our model can be implemented in state-of-the-art experiments. We suggest some experimental proposals: periodically driven on-site energies can be realized in dissipative single-molecule junctions \cite{sowa2017environment} by applying AC bias voltage, or alternatively, in a network of laser-written waveguides \cite{biggerstaff2016enhancing} by a time-dependent refractive index using the Pockels effect. Time-dependent hopping terms can be implemented by varying the distance between the sites (to manipulate the dipolar interactions) in a chain of trapped-ions \cite{trautmann2018trapped} or Rydberg-dressed-atoms \cite{schempp2015correlated}. In the mentioned experiments, the dephasing rates have been simulated using controllable schemes.

Here we will show that in the presence of on-site and off-diagonal periodic driving, the efficiency is enhanced within a range of parameters including the amplitude, frequency, noise, and trapping rates. We refer to this phenomenon as driving-assisted open quantum transport (DAOQT), while in another regime, the driving plays a detrimental role for the efficiency. To this end, we solve the problem analytically, by utilizing an extension of the Floquet theorem to open systems in the Lindblad picture, as well as numerically. To find the transport efficiency, we find the approximate time-independent Markovian generator using Magnus expansion. The analytical results are compared to exact numerical results.

\section{\label{sec:model}MODELS}
\subsection{\label{sec:diag-drive}On-Site Driving}
\begin{figure}
\includegraphics[width=0.91\columnwidth]{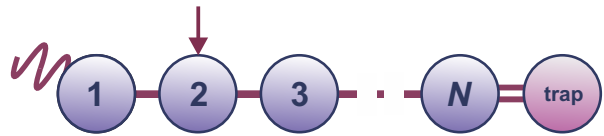}
\caption{\label{illustration}(Color online) Illustration of the network of qubits forming a linear chain. All sites are subject to dissipation and dephasing noises. On-site driving is applied to site 1 and off-diagonal driving is applied to all the couplings. Excitation initiates at site 2. The $N$-th site is irreversibly connected to a trap site.}
\end{figure}

We consider a network of $N$ coupled qubit sites that may support excitations and are subject to an external driving. First, we consider on-site periodic driving (see Fig. \ref{illustration}) with a period $T=2\pi/\Omega$. In a tight-binding approximation, the Hamiltonian is
\begin{eqnarray}
\nonumber H(t)&=&\sum_{k=1}^N \Big(\omega_k+\Delta_k \cos(\Omega t)\Big) \sigma_k^+ \sigma_k^- \\ &&+
 \sum_{k<l}\nu_{k,l} \Big(\sigma_k^+ \sigma_l^- +\text{H.c.}\Big),
\end{eqnarray}
where $\sigma_k^+ (\sigma_k^-)$ are the raising (lowering) operators for site $k$, $w_k$ is the site excitation energy, $\Delta_k$ is the magnitude of on-site driving, and $v_{k,l}$ is the hopping rate between the sites $k$ and $l$ whose values determine the topology of the network. Note that we have set $\hbar=1$. Here we consider the one-exciton manifold, which is  reasonable for low energy systems such as the light-harvesting complexes \cite{engel2007evidence,duan2017nature}. In a one-exciton manifold one replaces the lowering operator $\sigma_k^- \rightarrow |0\rangle\langle k|$, which is a projection operator from a localized excitation at site $k$ to the vacuum $|0\rangle$. In this manner, the Hamiltonian reads
\begin{eqnarray}\label{hamilton_onsite}
\nonumber H(t)&=&\sum_{k=1}^N \Big(\omega_k+\Delta_k \cos(\Omega t)\Big) |k\rangle\langle k| \\ &&+ \sum_{k<l}\nu_{k,l} \Big(|k\rangle\langle l| +\text{H.c.}\Big).
\end{eqnarray}
Our model without the driving term is comparable to those in Refs. \cite{plenio2008dephasing,kassal2012environment}.

To study the transport efficiency mediated by the Hamiltonian, we introduce environmental effects modeled by two distinct types of Markovian noise processes of Lindblad type \cite{lindblad1976generators,gorini1976completely}. The first is a dissipative process at a rate $\mu_k$ that reduces the exciton population, which is described by the super-operator
\begin{equation}
\mathcal{D}_{\mathrm{diss}}\,\rho=\sum_{k=1}^N \mu_k\Big[-\{|k \rangle\langle k|,\rho\}+2|0 \rangle\langle k|\rho|k \rangle\langle 0|\Big],
\end{equation}
where $\{\cdot,\cdot\}$ is an anti-commutator. The second is population-conserving dephasing process (phase randomization due to vibrational modes of phonon bath) at a rate $\gamma_k$,
\begin{equation}
\mathcal{D}_{\mathrm{deph}}\,\rho=\sum_{k=1}^N \gamma_k\Big[-\{|k \rangle\langle k|,\rho\}+2|k \rangle\langle k|\rho|k \rangle\langle k|\Big].
\end{equation}
In the microscopic derivation, one derives the Born-Markov approximation using a specific spectral density (for instance in Ref. \cite{mohseni2008environment}) to get the master equation. The information about the bath temperature is contained in $\gamma_k$. To calculate the transport efficiency, we connect the site $m$ to a trap site, denoted by $|N+1\rangle$, at a rate $\kappa$. The excitation is transfered irreversibly from site $m$ to $N+1$, which is also described by a super-operator,
\begin{equation}
\mathcal{D}_{\mathrm{trap}}\,\rho=\kappa\Big[-\{|m\rangle\langle m|,\rho\}+2|N+1\rangle\langle m|\rho|m\rangle\langle N+1|\Big].
\end{equation}
The Hilbert space $\mathcal{H}$ is extended to contain the trap and the vacuum states:
\begin{equation}\label{hilbert}
\mathcal{H}=\mathcal{H}_{\mathrm{sites}}+\mathcal{H}_{\mathrm{trap}}+\mathcal{H}_{0}.
\end{equation}
Hence, the complete equation of motion is
\begin{equation}\label{lindblad}
\dot{\rho}(t) = \mathcal{L}(t)\rho(t) = -i[H(t),\rho(t)] + \sum_i \mathcal{D}_i \rho(t)
\end{equation}
where $i=\{\text{diss},\text{deph},\text{trap}\}$. Note that the Liouvillian $\mathcal{L}$ is time-dependent. The transport efficiency $\eta$ is defined as the long-time population of the trap site,
\begin{equation}\label{eta_def}
\eta = \lim_{t\rightarrow\infty}p_{N+1}(t).
\end{equation}
Likewise, the loss-probability is the long-time evolution of vacuum population $p_0(t)$. The completeness of $\mathcal{H}$ demands that  $\sum_{n=0}^{N+1}p_n(t)=1$ for all $t\geq 0$ and this may be used to check the consistency of later calculations.

\subsection{\label{sec:off-diag-drive}Off-Diagonal Driving}
The second interesting case is when the coupling rates vary in time. For example, in a network of trapped ions the coupling may be varied by altering the distance between the ions. The Hamiltonian is
\begin{eqnarray}
\nonumber H(t)&=&\sum_{k=1}^N \omega_k \sigma_k^+ \sigma_k^-  +
 \sum_{k<l}\nu_{k,l} \\ &&\times\Big(1+f_{k,l} \cos(\Omega t)\Big) \Big(\sigma_k^+ \sigma_l^- +\text{H.c.}\Big).
\end{eqnarray}
The dissipators are the same as in the on-site driving case.


Our study will focus on how periodic driving affects the transport efficiency in a linear qubit network. Here we take a common coupling $\nu_{k,l}=\nu\delta_{k+1,l}$ and set it as our energy scale. To find out the parameter region in which the driving increases the transport, we analytically calculate the efficiency in the case of a localized initial excitation. We compare the analytical calculations to exact numerical results in specific parameter values.

\section{THE FLOQUET-MAGNUS-MARKOV EXPANSION}
To analytically obtain the efficiency $\eta$ we need to find the asymptotic solution $\rho(\infty)$ for the corresponding master equation. In principle, one could do this by calculating $\rho(\infty)=\lim_{t\rightarrow\infty}\mathcal{T}\exp\Big(\int_0^t\mathrm{d}\tau\,\mathcal{L}(\tau)\Big)\rho(0)$ where $\mathcal{T}$ is the time-ordering product, but this approach is computationally cumbersome even for a small system. Instead, we solve the following steady-state equation: \cite{kassal2012environment},
\begin{equation}\label{solver}
\mathcal{L}^\epsilon\rho(\infty) = \lim_{\epsilon\rightarrow 0} \epsilon\rho(0),
\end{equation}
using Gaussian elimination to find $\eta$, defined in Eq. (\ref{eta_def}). Here $\mathcal{L}^\epsilon\rho=\mathcal{L}\rho+\epsilon p_{N+1}|N+1\rangle  \langle N+1|$. This is the situation where $\rho(0)$ is being injected at rate $\epsilon$, and the limit is used to avoid trivial solutions. An intricacy arises because in our case the superoperator $\mathcal{L}$ depends explicitly on time, and at $t\rightarrow\infty$ the value is indefinite in the presence of periodic driving. We then consider the stroboscopic evolution using the Floquet theorem, but in our case it should be extended for open systems described by time-periodic Linblad master equations. 

We adopt the method in Ref. \cite{dai2016floquet} which we refer to as Floquet-Magnus-Markov (FMM) expansion. We begin with a time-dependent Markovian master equation,
\begin{equation}
\dot{\rho}(t)=\mathcal{L}(t)\rho(t).
\end{equation}
We can formally write the solution as
\begin{equation}
\rho(t_f)=\mathcal{V}(t_f,t_i)\rho(t_i) 
\end{equation}
where the propagator $\mathcal{V}(t_f,t_i)$ satisfies the divisibility condition, $\mathcal{V}(t_f,t_i)=\mathcal{V}(t_f,t_0)\mathcal{V}(t_0,t_i)$, and takes the form of $\mathcal{V}(t_f,t_i)=\text{e}^{\mathcal{L}(t_f)\delta t_f}\textellipsis\text{e}^{\mathcal{L}(t_k)\delta t_k}\textellipsis\text{e}^{\mathcal{L}(t_i)\delta t_i}$. Recall that if $\mathcal{L}(t)$ is time periodic, the propagator is also periodic,
\begin{equation}
\mathcal{V}(t_f,t_i)=\mathcal{V}(t_f+T,t_i+T),
\end{equation}
and depends only on $t_f-t_i$. 
In the Floquet picture, any periodic evolution operator can be decomposed into two parts: one contains the time-independent effective Floquet Liouvillian $\mathcal{L}_\text{F}[t]$ which is a functional of a starting time $t$, and another is the periodic micromotion of the driven system. Thus we can divide the propagator into three parts containing the Floquet Liouvillian in the middle, and the micromotions to account for the evolution before and after the Floquet part,
\begin{eqnarray}\label{Floquet_micromotion}
\mathcal{V}(t_f,t_i)&=&\mathcal{V}(t_f,t_0+nT)\mathcal{V}(t_0+nT,t_0)\mathcal{V}(t_0,t_1) \nonumber \\
	&=& \mathcal{V}(t_f,t_0+nT)\text {e}^{n\mathcal{L}_\text {F}[t_0]T}\mathcal{V}(t_0,t_i) \nonumber \\
	&=& \mathcal{V}(t_f,t_0+nT)\text {e}^{-\mathcal{L}_\text {F}[t_0]\delta t_f}\text {e}^{-\mathcal{L}_\text {F}[t_0](t_f-t_i)} \nonumber \\
	&& \times \;\text {e}^{\mathcal{L}_\text {F}[t_0]\delta t_i}\mathcal{V}(t_0,t_i) \nonumber \\
	&=& \mathcal{K}(\delta t_f) \text {e}^{\mathcal{L}_\text {F}[t_0] (t_f-t_i)}\mathcal{J}(\delta t_i),
\end{eqnarray}
where $\mathcal{K}(t)=\mathcal{V}(t_0+t,t_0)\text {e}^{-\mathcal{L}_\text {F}[t_0]t}$ and $\mathcal{J}(t)=\text {e}^{\mathcal{L}_\text {F}[t_0]t}\mathcal{V}(t_0,t_0+t)$ are the ''kick'' superoperators describing the micromotions, $\delta t_f=t_f - (t_0 + nT)$, and $\delta t_i = t_i - t_0$. A different starting time $t_0$ corresponds to a different $\mathcal{V}(t_0+T,t_0)$. Without loss of generality, we set $t_i=t_0=0$ and $t_f=t$, resulting in $\mathcal K(\delta t_f)=\mathcal K(t-nT)$ and $\mathcal{J}(\delta t_i)=1$. From now on, we simply refer to $\mathcal{L}_\text F [t_0]$ as $\mathcal{L}_\text F$. Thus the propagator in Eq. (\ref{Floquet_micromotion}) becomes
\begin{equation}\label{propagator}
\mathcal{V}(t,0)=\mathcal{K}(t-nT)\text {e}^{\mathcal{L}_{\text F}t}.
\end{equation}

Now we can describe the dynamics of the driven open system stroboscopically. However, this method is generally available only at high driving frequency $\Omega$. In the calculation of transport efficiency $\eta$, one is interested in the asymptotic solution, $t\rightarrow\infty$, so that the micromotion Eq. (\ref{propagator}) is negligible,  $\mathcal{K}(t-nT)\approx 1$. Hence, for the asymptotic dynamics in Eq. (\ref{solver}) we can replace $\mathcal{L}$ with $\mathcal{L}_{\text F}$.

Having the asymptotic dynamics governed by the Floquet Liouvillian $\mathcal{L}_{\text F}$, now we want to derive its approximate form. To this end, we use Magnus expansion \cite{blanes2009magnus} for $\mathcal{L}_{\text F}=\mathcal{L}^{(0)}_{\text F}+\mathcal{L}^{(1)}_{\text F}+\mathcal{L}^{(2)}_{\text F}+\textellipsis$. The first three terms are
\begin{eqnarray}
\mathcal{L}_\text{F}^{(0)} =&& \frac{1}{T}\int_0^T {d}t\; \mathcal{L}(t) , \nonumber \\
\mathcal{L}_\text{F}^{(1)} =&& \frac{1}{2T}\int_0^T {d}t_1\int_0^{t_1} {d}t_2\; [\mathcal{L}(t_1),\mathcal{L}(t_2)], \nonumber \\
\mathcal{L}_\text{F}^{(2)} =&& \frac{1}{6T}\int_0^T {d}t_1\int_0^{t_1} {d}t_2\int_0^{t_2} {d}t_3\;
\Big(\,[\mathcal{L}(t_1),[\mathcal{L}(t_2),\mathcal{L}(t_3)]] \nonumber \\ &&+\;[[\mathcal{L}(t_1),\mathcal{L}(t_2)],\mathcal{L}(t_3)]\,\Big). \label{floquet-magnus}
\end{eqnarray}

To apply this method, we cast the system Hamiltonian into the following form,
\begin{equation}\label{H_cast}
H(t) = H_0 + H_1 F(t)
\end{equation}
where $F(t)$ is the time-dependent part of $H(t)$. In our case  $F(t)=\cos(\Omega t)$. Using the FMM expansion, we work out the first three leading terms of $\mathcal L_\text {F}$,
\begin{eqnarray}
\mathcal{L}_\text{F}^{(0)} \rho = -i[H_0,\rho] + \sum_i\mathcal{D}_i \rho, \label{L0} \\
\mathcal{L}_\text{F}^{(1)} \rho = 0,\qquad\qquad \label{L1}
\end{eqnarray}
and, after a tedious calculation, the second order is 
\begin{widetext}
\begin{eqnarray}
\mathcal{L}_\text{F}^{(2)} \rho = &-&\frac{\text{i}}{\Omega^2}[H_1,\sum_i\sum_j\mathcal{D}_i\mathcal{D}_j\rho] +\frac{2\text{i}}{\Omega^2}\sum_i\mathcal{D}_i[H_1,\sum_j\mathcal{D}_j\rho] -\frac{\text{i}}{\Omega^2}\sum_i\sum_j\mathcal{D}_i\mathcal{D}_j[H_1,\rho] +\frac{2}{\Omega^2}[H_0,[H_1,\sum_i\mathcal{D}_i\rho]] \nonumber \\ &-&\frac{1}{\Omega^2}[H_1,[H_0,\sum_i\mathcal{D}_i\rho]] -\frac{1}{4\Omega^2}[H_1,[H_1,\sum_i\mathcal{D}_i\rho]] -\frac{1}{\Omega^2}[H_0,\sum_i\mathcal{D}_i[H_1,\rho]] -\frac{1}{\Omega^2}[H_1,\sum_i\mathcal{D}_i[H_0,\rho]] \nonumber \\ &+&\frac{1}{2\Omega^2}[H_1,\sum_i\mathcal{D}_i[H_1,\rho]] -\frac{1}{\Omega^2}\sum_i\mathcal{D}_i[H_0,[H_1,\rho]] +\frac{2}{\Omega^2}\sum_i\mathcal{D}_i[H_1,[H_0,\rho]] -\frac{1}{4\Omega^2}\sum_i\mathcal{D}_i[H_1,[H_1,\rho]], \label{L2}
\end{eqnarray}
\end{widetext}
and $\mathcal L_\text F = \mathcal L_\text F ^{(0)} + \mathcal L_\text F ^{(1)} + \mathcal L_\text F ^{(2)} + O(1/\Omega^3)$. In this paper, we work in the fast driving regime $\Omega>\nu_{k,l}$, in which the analytical FMM expansion is plausible. However, in the next section we show that maximum efficiency enhancement is achieved at $\Omega$ slower but near the coupling rate $\nu_{k,l}$. 

It should be noted that Eq. (\ref{L2}) is not in Lindblad form and thus the FMM expansion is not completely positive (CP). Instead, only the zeroth order is CP, and the rest is only approximately CP as one considers finite expansion terms.

Numerical results are done by calculation of the exact dynamics using master equation solver in the \verb+QuTiP+ package \cite{johansson2013qutip}.

\section{RESULTS AND DISCUSSION}
\vbox{In order to obtain results which are not blurred by the network complexity, we consider a short linear chain where $N=3$ and the rates $\mu,\gamma$ are equal on all sites. It is shown in Ref. \cite{kassal2012environment} that ENAQT persists in an ordered system, in contrast to previous studies \cite{mohseni2008environment,plenio2008dephasing,rebentrost2009environment} where previously it was thought that the ENAQT would be possible only in a disordered system due to interplay between Anderson localization \cite{anderson1958absence} and dephasing noise. In fact, in an ordered system, the only case where ENAQT is impossible is in end-to-end transport. To this end, we consider an ordered system in which we renormalize $\omega_k=0$. The renormalization does not break the validity of the local master equation Eq. (\ref{lindblad}) as long as $\omega\gg\nu$ \cite{hofer2017markovian}. The excitation initiates at site $i=2$, $\rho(0)=|2\rangle\langle 2|$, and the trap is connected to site $m=3$.}

\subsection{On-Site Driving}
We begin with a brief analysis of the transport where the periodic driving is applied equally on all the sites, $\Delta_{k}(t)=\Delta$. In this case there is no enhancement to $\eta$ and, in fact, there is no effect to the transport at all. One can check in an ordered system that $\eta$ is independent of $\omega$; see Eq. (\ref{eta_benchmark}). Thus a homogenous driving has no effect to $\eta$ since it keeps the system ordered in a rotating frame. Here we set $\Delta_k$ nonzero only for  site $k=1$. 

In the following we will apply the FMM expansion to find $\eta$ via the infinite-time stroboscopic evolution. To this end, we separate the time-dependence of the Hamiltonian in the form of Eq. (\ref{H_cast}), where
\begin{equation}\label{H0H1onsite}
H_0 = \nu\sum_{k<l}^{N=3} \Big(|k\rangle\langle l| +|l\rangle\langle k|\Big),\quad H_1 = \Delta|1\rangle\langle1|,
\end{equation}
and $F(t)=\cos(\Omega t)$. Inserting Eq. (\ref{H0H1onsite}) into Eqs. (\ref{L0})--(\ref{L2}), one obtains a set of linear differential equations written in Appendix \ref{app:onsite}. Transport efficiency is obtained by solving Eq. (\ref{solver}). The solution is found using Gaussian elimination, resulting in $\eta$ in the form of a rational function,
\begin{equation}\label{eta}
\eta = \kappa\nu^2\frac{\sum\limits_{n=0}^5 A_{n}(\nu,\gamma,\mu,\kappa,\Omega)\Delta^{2n} }{\sum\limits_{n=0}^6 B_{n}(\nu,\gamma,\mu,\kappa,\Omega)\Delta^{2n}},
\end{equation}
with the coefficients $A_{n}$ and $B_{n}$ written in the Supplementary Material [link will be inserted by the publisher]. The solution without the presence of driving, $\eta_0$, is obtained by putting $\Delta=0$, or equivalently by taking the limit $\Omega\rightarrow\infty$ so that $\cos(\Omega t)$ averages out to zero,
\begin{equation}\label{eta_benchmark}
\eta_0 = \kappa\nu^2\frac{ \alpha_2\gamma^2+\alpha_1\gamma+\alpha_0}{\beta_3\gamma^3+\beta_2\gamma^2+\beta_1\gamma+\beta_0}
\end{equation}
where
\begin{eqnarray} 
\alpha_0&=&(\kappa + 2 \mu) (\nu^2 + 2 \mu^2), \nonumber \\
\alpha_1&=&2\gamma[\mu(\kappa+4\mu)+\nu^2], \nonumber \\ 
\alpha_2&=&4\mu, \nonumber \\
\beta_0&=&[(\kappa + 2 \mu)\nu^2 + \mu^2(\kappa + \mu)] \nonumber \\ && \times [2 m (\kappa + 2 \mu)^2 + (\kappa + 4 \mu) \nu^2], \nonumber \\
\beta_1 &=& 2[\mu^2 (\kappa + \mu) (\kappa + 2 \mu) (\kappa + 6 \mu)  \nonumber\\&& +
 \mu (5 \kappa^2 + 20 \kappa \mu + 18 \mu^2) \nu^2 + (\kappa + 3 \mu) \nu^4], \nonumber \\
 \beta_2&=&4\mu [2 \kappa \mu (\kappa + \mu) + 6 \mu^2 (\kappa + m) + 3 \kappa \nu^2 + 4 \mu \nu^2], \nonumber \\
  \beta_3&=&8\mu^2 (\kappa + \mu).
\end{eqnarray}
This expression matches with the results in Refs. \cite{kassal2012environment,plenio2008dephasing}. ENAQT is defined as the difference between the maximum efficiency ($\eta_{0\text {max}}$), where the dephasing is optimum, $\gamma_{\text{opt}}$, and the efficiency without driving ($\eta_0$) \cite{kassal2012environment}. The existence of $\gamma_{\text{opt}}$ is possible whenever $\partial\eta_0/\partial\gamma = 0$, which is dominantly occurring for small $\mu$ and $\kappa$  in the order of magnitude up to $O(10^{-1}\nu)$. 

\begin{figure}
\includegraphics[width=\columnwidth]{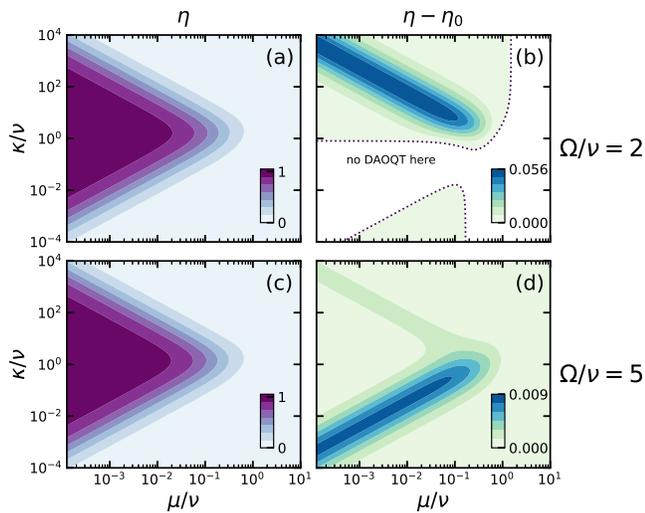}
\caption{\label{mk_contour} (Color online) Analytical characterization of efficiency of on-site DAOQT as a function of loss ($\mu$) and trapping ($\kappa
$) rates in a $N=3$ system with $\Delta=2\nu$ and $\gamma=0$ (no dephasing). The efficiency $\eta$ is computed using second order FMM expansion whose result is in Eq. (\ref{eta}), and $\eta_0$ is from Eq. (\ref{eta_benchmark}). (a) Efficiency for $\Omega=2\nu$ and (b) the corresponding DAOQT defined as the difference between $\eta$ and $\eta_0$. The white area between the two dotted lines indicates the region where DAOQT is not possible (see text). (c) Efficiency for $\Omega=5\nu$ and (d) the corresponding DAOQT; here DAOQT occurs dominantly at a low $\kappa$ regime, in contrast to the $\Omega=2\nu$ case. }
\end{figure}

To understand the role of the driving to the transport, we first analyze $\eta$ at certain values of $\Delta$ and $\Omega$. We refer to the efficiency enhancement $\eta-\eta_0$ due to  driving as the DAOQT where $\eta_0$ is the efficiency for $\Delta=0$ (or $\Omega\rightarrow\infty$ when the driving averages out). Figures \ref{mk_contour}(a) and \ref{mk_contour}(c) shows $\eta$ as a function of $\mu$ and $\kappa$ in the $N=3$ chain with the trap site at one end, the driven site at the other end, and the initial site in the middle. As one may expect, in general high efficiency is achieved at small dissipation $\mu/\nu$. The symmetric triangle-like profile corresponding to high $\eta$ in Figs. \ref{mk_contour}(a) and \ref{mk_contour}(c) is characteristic for ENAQT in ordered systems \cite{kassal2012environment}. Several limiting cases can immediately be obtained, regardless of the number of sites: (1) at $\kappa\ll\mu$, the excitation is easily lost before it is efficiently trapped, (2) at $\kappa\gg\mu$, the suppression of transport is due to the quantum Zeno effect, in which the coherence vanishes due to rapidly measurement done by the sink, (3) at $\Delta>0$ the triangle center is shifted away from $\kappa=\sqrt{2}\nu$. Thus the driving opens some regimes which were previously not available for an efficient transport.

The DAOQT is apparent in Figs. \ref{mk_contour}(b) and \ref{mk_contour}(d) where the white area below the dotted line indicates $\eta-\eta_0<0$ (no enhancement). According to the FMM expansion, driving affects the transport as a second order process proportional to $\Omega^{-2}$; see Eq. (\ref{L2}). Thus we expect DAOQT to occur significantly at low $\Omega/\nu$. We first take $\gamma=0$ to isolate the effect from the enhancement due to ENAQT. For small dissipation rates, the transport regimes depend on the competition between $\kappa$ and $\Delta\nu^2/\Omega^2$. Positive DAOQT is achieved in $\kappa$ regimes dependent of $\Omega/\nu$. In Fig. \ref{mk_contour}(b), where $\Omega=2\nu$, DAOQT occurs dominantly at the quantum Zeno regime with $\kappa\gg\nu$. In contrast, in Fig. \ref{mk_contour}(d) DAOQT occurs dominantly at slow trapping rates, although the enhancement is small for sufficiently large $\Omega/\nu$.

\begin{figure}
\includegraphics[width=\columnwidth]{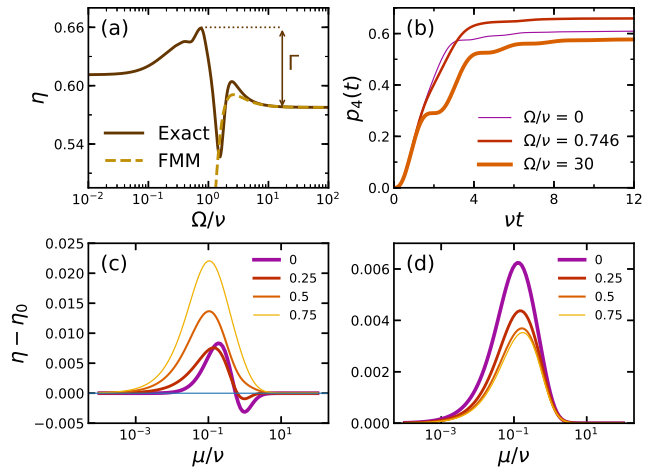}
\caption{\label{fig_onsite} (Color online) (a) Transport efficiency $\eta$ as a function of driving frequency for $\Delta=2\nu$, $\gamma=0$, $\mu=0.1\nu$, and $\kappa=0.8\nu$ [corresponds to high DAOQT region in Fig. \ref{mk_contour}(a)]. The FMM approximation is accurate in high driving frequencies. DAOQT is maximized at $\Omega=0.746\nu$ with  enhancement $\Gamma=8.14\%$ and the global minimum is at $\Omega=1.56\nu$. (b) Exact time evolution for trap site population with the parameters from (a), $\Omega=0$ is the case of static disorder, $\Omega=0.746\nu$ is the optimum, and $\Omega=30\nu$ is the high frequency limit. Effect of increasing dephasing $\gamma$ (analytical) is shown in (c) for $\Omega=2\nu$, where it contains negative enhancement for $\gamma=0$, and (d) $\Omega=5\nu$. The thinner line indicates higher value of $\gamma$.}
\end{figure}

In Fig. \ref{fig_onsite}(a), we compare the efficiency from exact numerical result and second order FMM expansion. The analytical calculation is accurate for high $\Omega/\nu$ and captures the local maximum around $\Omega=2\nu$ although not exactly. At small $\Omega/\nu$, the FMM does not capture the interesting global maximum. The validity of second-order FMM expansion, according to $\mathcal{L}^{(2)}_\mathrm{F}$ in Eq. (\ref{L2}), relies on the values of ${\kappa^2\mu}/{\Omega^2\nu}$, ${\Delta\gamma\mu}/{\Omega^2\nu}$, and such coefficients, being small [see Eq. (\ref{ODE_onsite}) for on-site driving]. In short-time evolution of $\rho(t)$, the approximation still mimics the exact dynamics stroboscopically, but for long-time evolution it may lead to a different steady state---although for the parameters in Fig. \ref{fig_onsite}(a) the difference is relatively small for $\Omega/\nu\gtrsim2$. It is shown that the FMM expansion clearly deviates from the exact result for $\Omega<\nu$. Nevertheless, it reproduces the local peak at around $\Omega=2\nu$. For frequencies larger than this the FMM method is reliable. The efficiency contour in Fig. \ref{mk_contour} takes the value of $\Omega=2\nu$, which corresponds to the local maximum that is adequately reproduced by the FMM expansion, and $\Omega=5\nu$, which falls in the range where the FMM method is reliable.

Here, we define the maximum DAOQT as
\begin{equation}
\Gamma(\gamma,\mu,\kappa,\Omega) = \eta_{\text {max}}(\gamma,\mu,\kappa,\Omega) - \lim_{\Omega\rightarrow \infty}\eta(\gamma,\mu,\kappa,\Omega).
\end{equation}
where the maximizing frequency is $\Omega_\text {opt}$. The $\Omega\rightarrow\infty$ case is when the driving is averaged out and the efficiency draws back to $\eta_0$. On the other hand, $\eta(\Omega=0)$ corresponds to the existence of a static disorder at site 1---though this is not always the case with different regimes of $\mu$ and $\kappa$. The maximum efficiency $\eta_\text{max}$ is larger than $\eta(\Omega=0)$, indicating that the existence of driving may enhance the transport further than ENAQT in a disordered chain. On the other hand, we can observe that the driving does see a negative enhancement near $\Omega=\nu$. 

\begin{figure}
\includegraphics[width=0.8\columnwidth]{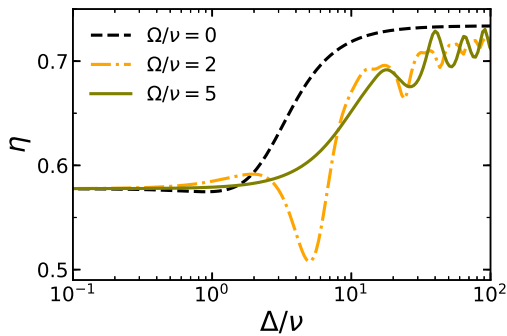}
\caption{\label{f_omega025} (Color online) Exact transport frequency as a function of driving amplitude with paramaters as in Fig. \ref{fig_onsite}(a). In the high amplitude limit, $\eta$ converges to $0.73$ and is limited by the static disordered case ($\Omega/\nu=0$). At nonzero driving frequencies, higher $\Delta/\nu$ does not always guarantee higher efficiency than the static disordered chain.}
\end{figure}

The DAOQT is also interrelated with quantum coherence between the sites. Figure \ref{fig_onsite}(b) shows that in low frequencies ($\Omega<\nu$) the coherent evolution, indicated by wiggly lines in the population dynamics, in the trap site is suppressed, while in $\Omega=5\nu$ it is retained. In this case, the driving may have a detrimental effect on the coherence. Nevertheless, efficient transport due to dephasing noise (ENAQT) also destroys quantum coherence \cite{plenio2008dephasing}. This is because coherence increases the time needed by the excitation to wander around the chain, resulting in less efficient transport.

In general the behavior of the driven system under a dephasing noise is similar to ENAQT for a disordered chain. DAOQT tend to arise for $\gamma>0$ if $\mu$ and $\kappa$ falls in the ENAQT regime. Onsite driving creates a dynamical disorder that is time periodic, combined together with $\gamma$, they suppress the Anderson localization and increases the directivity of the transport. The subtleties arise from the fact that the driving frequency $\Omega$ may not always create a constructive interference when it competes with $\mu$ and $\kappa$. In Fig. \ref{fig_onsite}(c)and \ref{fig_onsite}(d), we compare the DAOQT as function of $\mu$ in the presence of dephasing $\gamma$. Note that for $\Omega=2\nu$ and $\gamma=\nu$ the transport is slightly suppressed for some values of $\mu$. For $\Omega=5\nu$, where the DAOQT is comparatively small, $\gamma>0$ suppresses a transport. However, in this case the difference is negligible and the driving already starts to average out.

The underlying mechanism of the on-site DAOQT lies at the formation of periodic dynamic disorder at the driven site. By periodically altering the energy level of that site, one can increase (or decrease) the directivity of the transport by controlling the interference effects at the sites. If the periodic driving can direct the excitation to site 3 (connected to the trap) at a right period such that the excitation remains at site 3 for a long time, there will be a gain in efficiency. The range of $\Omega$ that brings the benefit is of course depends on the coupling, trapping, and dissipative rates. In general, the maximum enhancement $\Gamma$ is found to be near $\Omega=\nu$. Note that the periodic driving of site energy may yield a higher efficiency enhancement than in the case of static disorder. This is because a static disorder cannot provide a dynamical control to alter the interplay between coherence and dissipation, i.e., to make the excitation stays longer at a site connected to the trap---this is the key mechanism of DAOQT. We term the elongation of a visit by the excitation ''population congestion''. 

Figure \ref{f_omega025} shows the effect of driving strength to efficiency. $\Omega=0$ (dashed line) corresponds to static disorder, in comparison to the dynamic ones. At $\Delta\gtrsim\nu)$, the driven transport (for $\Omega=2\nu$ and $\Omega=5\nu$) is more effecient than the static one. This shows that periodic driving enhances the transport even in the presence of static disorder, while, at large $\Delta/\nu$, the periodic driving is not beneficial to the transport compared to static disorder, and at $\Delta>100\nu$ the efficiency stops gaining as the site energy becomes too high for the excitation to enter. Note that the global minimum for $\Omega=2\nu$ exists because in the population congestion mechanism tends to direct the excitation to the wrong site, i.e., to site 1 instead of site 3, which is connected to the trap. Negative enhancement also occurs in Fig. \ref{fig_onsite}(a) for $\Omega$ slightly larger than $\nu$. We will find again the negative enhancement in off-diagonal (coupling) driving with different congestion dynamics.

\subsection{Off-Diagonal Driving}
We implement a homogeneous driving, $f_{k,l}(t)=f$, to all the site couplings in $N=3$ ordered linear chain. The trap site is still connected to site 3, and the excitation also initiates at site 2. The corresponding Hamiltonian in the form of Eq. (\ref{H_cast}) is
\begin{equation}
H_0 = \frac{H_1}{f} = \nu\sum_{k<l}^{N=3} \Big(|k\rangle\langle l| +|l\rangle\langle k|\Big)
\end{equation}
and $F(t)=\cos(\Omega t)$. The procedures are the same as in the on-site driving and the corresponding dynamical equations are written in Appendix \ref{app:offdiag}. In the limits of $f=0$ or $\Omega\rightarrow\infty$, we also obtain $\eta_0$ as in Eq. (\ref{eta_benchmark}).

\begin{figure}
\includegraphics[width=\columnwidth]{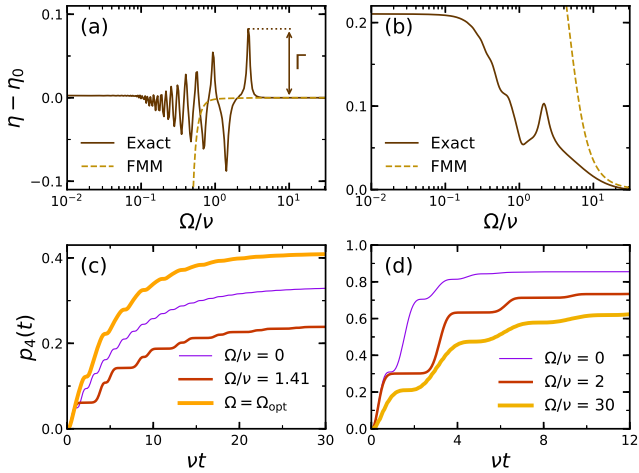}
\caption{\label{fig_offdiag}(Color online) Characterization of off-diagonal DAOQT. (a) The DAOQT for $f=1$, $\gamma=0$, $\mu=0.05\nu$, and $\kappa=0.1\nu$ (slow trapping). Maximum DAOQT is at $\Omega_\text {opt}=2.8\nu$ with $\Gamma=8.25\%$ while the global minimum is at $\Omega=1.41\nu$ and $\eta_0=0.33$. (b) Same parameters as in (a) but with $\kappa=5\nu$ (fast trapping), here $\eta_0=0.64$. The behavior is totally different because there is no oscillating pattern such as in (a). The exact time evolution for trap site population is shown in (c) for the slow trapping and (d) fast trapping.}
\end{figure}

The off-diagonal DAOQT characteristics are shown in Fig. \ref{fig_offdiag}, where we plot $\eta-\eta_0$ for slow and fast trapping in Figs. \ref{fig_offdiag}(a) and \ref{fig_offdiag}(b), respectively. Fast trapping is the case for $\kappa$ in the quantum Zeno regime. Within this regime, DAOQT ceases to exist and $\Omega=0$ is the most efficient case for transport. Again, the FMM expansion is accurate only with large $\Omega/\nu$ as we have discussed in on-site driving. However, in this case it does not capture the maxima as it does before. For $f=1$, $\Omega=0$ means that the coupling strength is doubled to $2\nu$. This doubling has little or no effect for the slow trapping but is prominent for the fast trapping due to the dominance of the incoherent population hopping over coherent oscillation. One can observe an interesting oscillation pattern with sharp peaks and troughs in Fig. \ref{fig_offdiag}(a), which indicates that the system is sensitive to external driving only between $\Omega= 10^{-1}\nu$ and $3\nu$. This is in contrast with the onsite driving in Fig. \ref{fig_onsite}(a) where the driving averages out smoothly. Here the driving appears to be suddenly averaged out after the global maximum in $\Omega\approx 3\nu$. We will discuss this pecular behavior when we point out the DAOQT mechanism for off-diagonal driving below.

\begin{figure}
\includegraphics[width=\columnwidth]{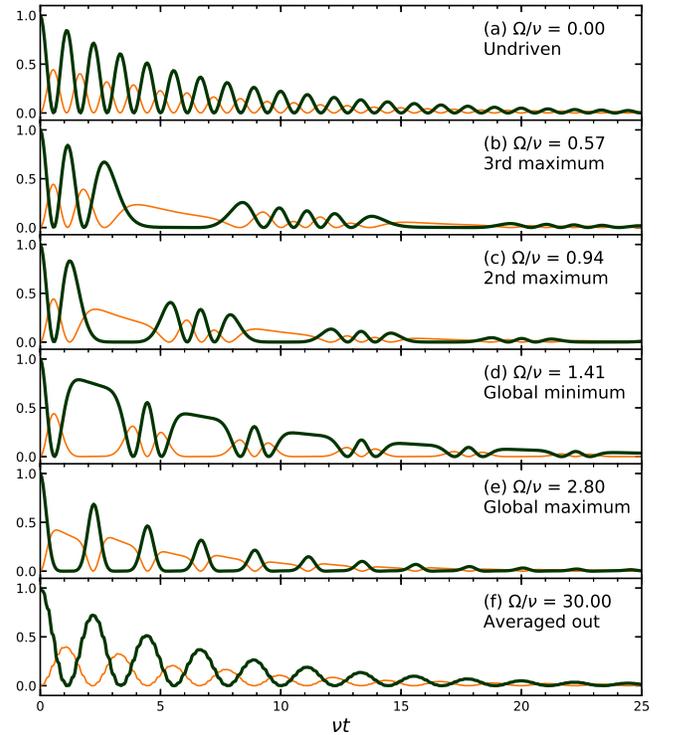}
\caption{\label{coh_offdiag} (Color online) Exact population dynamics for initial site population, $p_2(t)$ (thick lines), and site 3 population, $p_3(t)$ (thin lines), with the frequencies (a)--(f) correspond to the interesting values in Fig. \ref{fig_offdiag}(a) with the same parameters.}
\end{figure}

Figures \ref{fig_offdiag}(c) and \ref{fig_offdiag}(d) show the time evolutions for the trap site corresponding to the slow and fast trapping, respectively. It appears at first sight that the sign of coherent oscillation (wiggly lines) exists even for fast trapping, whereas incoherent transport is expected due to the quantum Zeno effect. Instead, this is because the sites are periodically uncoupled---observe that, at $t_n=(2n+1)\pi/\Omega$ with $n\in\mathbb{Z}$, $H(t_n)=0$. At these times, the transport is governed only by dissipation and trapping, while the evolution is incoherent. This periodic freezing of the system turns out to be the key of the DAOQT mechanism for off-diagonal driving which we will explain below.

The DAOQT mechanism is different with the one in on-site driving case since in this case the driving does not create energy disorder. It is purely the interplay between periodic coupling and noise rates. We start with the undriven chain, $\Omega=0$, where the sites coupling strength is effectively $2\nu$. The diagonalized Hamiltonian has eigenfrequencies of  $\lambda=\{0,\pm 2\sqrt{2}\nu\}$, and correspondingly it can be seen in Fig. \ref{coh_offdiag}(a) that the populations $p_2$ and $p_3$ oscillates with a period of $\pi/2\sqrt{2}\nu$ up to an exponential decay with the rate $2\mu$ for $p_2$ and $2(\mu+\kappa)$ for $p_3$. When the driving is nearly averaged out, Fig. \ref{coh_offdiag}(f), the frequency sets back to $\sqrt{2}\nu$. These two cases portray the typical dynamics in ENAQT. Figures \ref{coh_offdiag}(b),\ref{coh_offdiag}(c), and \ref{coh_offdiag}(e) shows the dynamics when the system takes benefit from driving (the values of $\Omega/\nu$ correspond to maxima of Fig. \ref{fig_offdiag}(a)). At these frequencies, it is apparent that the excitation is congested at site-3 and completely vanishes from site-2 (while the other half of the population is at site-1). The complete suppression of $p_2$ indicates that at these frequencies the population congestion mechanism should give a maximum benefit to the transport, i.e., resulting in the local maxima. We can observe that, at the frequencies corresponding to local maxima, the populations evolve with a pattern: for the global maximum the population $p_2(t)$ oscillates with one peak, for the second maximum with three peaks, and for the $n+1$-th maximum with $2n+1$ peaks. The driven dynamics with one peak, Fig. \ref{coh_offdiag}(e), gives the longest time possible for the excitation being in site 3, thus letting it be transferred to the trap at most. In contrast, in Fig. \ref{coh_offdiag}(d) the driving is congesting the population at the wrong site, resulting in negative enhancement. The mentioned mechanism is responsible for the existence of the sharp peaks and the apparently sudden averaging of Fig. \ref{fig_offdiag}(a). In fact, shortly after the global maximum, $\Omega\gtrsim 2.8\nu$, the driving is far from being averaged [compare Figs. \ref{coh_offdiag}(e) and \ref{coh_offdiag}(f)]. It appears so because the minimum number of peaks in the population dynamics is already reached.

Numerical studies indicate that $\mu$ and $\gamma$ are not dominant in determining the position of the global maximum and the spacing between the peaks in Fig. \ref{coh_offdiag}(a)---although the oscillating pattern still terminates shortly after the global maximum. This applies for slow trapping rates. Meanwhile $\mu$ does not affect the dynamics in the second order; see the corresponding dynamical equations in Appendix \ref{app:offdiag}.


\section{CONCLUSIONS}
We have discussed the transport characteristics of an exciton in a dissipative qubit network under on-site and off-diagonal periodic drivings. As a clear example, we have considered a linear ordered chain with $N=3$ consisting of an initial site, a driven site and a trap-connected site at opposite sides of the chain. We have shown that external periodic driving may increase the transport efficiency at driving frequencies near the coupling rate for on-site and off-diagonal driving. The maximum enhancement of on-site DAOQT occurs at frequencies $\Omega$ just below $\nu$. However, at other dissipative and trapping rates ($\mu$ and $\kappa$) we find that there is no enhancement since the static disorder case has the highest efficiency. On the contrary, at $\Omega\gtrsim\nu$ the efficiency may be suppressed lower than the undriven case. 

In general, the on-site driving opens the regimes which are previously not available for efficient transport. In the presence of off-diagonal driving, we found that the DAOQT shows an oscillatory pattern at low trapping rates and ceases to exist at high trapping rates due to the quantum Zeno effect. We proposed the DAOQT mechanism termed population congestion for both cases. For on-site driving, the efficiency enhancement is due to the formation of periodic dynamic disorder at the driven site. Periodic alteration of the energy level controls the interference effect at the sites, which in turn may increase or decrease the directivity of the transport depending on at which site the driving is directing the transport. For off-diagonal driving, the enhancement is the result of interplay between periodic coupling and noise rates.

\begin{acknowledgments}
F.P.Z. thanks Ministry of Higher Education and Research of Indonesia for Research Funding Desentralisasi 2019. The numerical results was obtained using code written in \verb+NumPy+ \cite{van2011numpy} and \verb+QuTiP+ \cite{johansson2013qutip}, and the figures were made using \verb+matplotlib+ \cite{hunter2007matplotlib}. 
\end{acknowledgments}

\appendix

\section{Analytical calculation of transport efficiency}
\subsection{\label{app:onsite}On-Site Driving}\label{sec:App_onsite}
The efficiency is calculated by finding the asymptotic solution using Eq. (\ref{solver}) with $\mathcal{L}=\mathcal{L}_\text{F}$ (FMM expansion to the second order) and initial condition $\rho(0)=|2\rangle\langle 2|$. Each of the component $\mathcal{L}_\text{F}\rho_{ij}$ consists of a linear equation. The corresponding equations for on-site driving are written below, and is solved via Gaussian elimination.
\begin{eqnarray}\label{ODE_onsite}
\mathcal{L}_\text{F}\rho_{11} &=& -2\mu\rho_{11}-2\nu\Im\rho_{12} \nonumber \\ &&+\frac{f\nu}{2\Omega^2}(\Delta\Im\rho_{12}+8\gamma\Re\rho_{12}+4\nu\Im\rho_{13}),\qquad\qquad\quad \nonumber \\
\mathcal{L}_\text{F}\rho_{22} &=&-2\mu\rho_{22}+2\nu\Im(\rho_{12}-\rho_{23}) \nonumber \\ &&- \frac{\Delta\nu} {2\Omega^2}(\Delta\Im\rho_{12}+8\gamma\Re\rho_{12}), \nonumber \\
\mathcal{L}_\text{F}\rho_{33} &=&-2(\mu+\kappa)\rho_{33}+2\nu\Im\rho_{23}-\frac{2\Delta\nu^2}{\Omega^2}\Im\rho_{13}, \nonumber \\
\mathcal{L}_\text{F}\rho_{44} &=& 2\kappa\rho_{33}, \nonumber \\
\mathcal{L}_\text{F}\rho_{12} &=&-2(\gamma+\mu)\rho_{12}+\text {i}\nu(\rho_{11}-\rho_{22}+\rho_{13}) \nonumber \\ &&+ \frac{\text i \Delta\nu}{2\Omega^2}\left(\nu(4\rho_{12}+\rho_{23}^*)-\frac{8\text {i}\gamma+\Delta}{4}(\rho_{11}-\rho_{22})\right), \nonumber \\
\mathcal{L}_\text{F}\rho_{13} &=& -(2(\gamma+\mu)+\kappa)\rho_{13}+\text {i}\nu(\rho_{12}-\rho_{23}) \nonumber \\ &&+ \frac{\text {i}\Delta\nu}{4\Omega^2}(\Delta\rho_{23}+4\nu(\rho_{33}-\rho_{11}+2\rho_{13})), \nonumber \\
\mathcal{L}_\text{F}\rho_{23} &=& -(2(\gamma+\mu)+\kappa)\rho_{23}-\text {i}\nu(\rho_{33}-\rho_{22}+\rho_{13}) \nonumber \\ &&+ \frac{\text {i}\Delta\nu}{4\Omega^2}(f\rho_{13}-4\nu(\rho_{12}^*+2\rho_{23})). 
\end{eqnarray}

\subsection{\label{app:offdiag}Off-Diagonal Driving}
The corresponding linear equations from Eq. (\ref{solver}) for off-diagonal driving are written below.  The solution $\eta$ for this case is also found using Gaussian elimination (not shown in this paper).
\begin{eqnarray}\label{ODE_off_diag}
\mathcal{L}_\text{F}\rho_{11} &=& -2\mu\rho_{11}-2\nu\Im\rho_{12} +\frac{f\nu}{\Omega^2}\Big(-8\gamma^2\Im\rho_{12} \nonumber \\ && +\frac{\nu}{2}(f-4)[\kappa\Re\rho_{13}-2\gamma(2\rho_{11}-2\rho_{22}+\Re\rho_{23})]\Big), \nonumber \\
\mathcal{L}_\text{F}\rho_{22} &=&-2\mu\rho_{22}+2\nu\Im(\rho_{12}-\rho_{23})+ \frac{f\nu} {\Omega^2}\Big(8\gamma^2\Im\rho_{12} \nonumber \\ &&-2(2\gamma+\kappa)^2\Im\rho_{23}+\nu(f-4)[-\kappa\Re\rho_{22}\nonumber \\
&&+2\gamma(\rho_{11} -2\rho_{22}+\rho_{33}+\Re\rho_{13})]\Big), \nonumber \end{eqnarray}

\begin{eqnarray}
\mathcal{L}_\mathrm{F}\rho_{33} &=&-2(\mu+\kappa)\rho_{33}+2\nu\Im\rho_{23}\nonumber \\
&&-\frac{f\nu}{\Omega^2}\Big(2(-2\gamma+\kappa)^2\Im\rho_{23}  +\frac{\nu}{2}(f-4)[4\gamma\rho_{22}\nonumber \\
&&+(\kappa-2\gamma)(2\rho_{33}+\Re\rho_{13})]\Big), \nonumber \\
\mathcal{L}_\text{F}\rho_{44} &=& 2\kappa\rho_{33}-\frac{f\nu\kappa}{\Omega^2}\Big(\kappa^2\Im\rho_{13}\nonumber \\
&&-\frac{\kappa}{4}(f-4)(\Re\rho_{23}-2\Re\rho_{12})\Big), \nonumber
\\
\mathcal{L}_\text{F}\rho_{12} &=&-2(\gamma+\mu)\rho_{12}+\text {i}\nu(\rho_{11}-\rho_{22}+\rho_{13})\nonumber \\
&&+ \frac{\text i f\nu}{4\Omega^2}\Big(4[\kappa^2\rho_{13}+4\gamma^2(\rho_{11}-\rho_{22})]\nonumber \\
&&-\text i \nu(f-4)[8\gamma(2\rho_{12}-\rho_{23})+\kappa(\rho_{23}^*-2\rho_{12})\nonumber \\
&&+8\gamma\Re(\rho_{23}-2\rho_{12})]\Big), \nonumber \end{eqnarray}

\begin{eqnarray}
\mathcal{L}_\text{F}\rho_{13} &=& -(2(\gamma+\mu)+\kappa)\rho_{13}+\text {i}\nu(\rho_{12}-\rho_{23})\nonumber \\
&&+ \frac{\text i f\nu}{4\Omega^2}\Big(\kappa^2\rho_{12} -\frac{\text i\nu}{4}(f-4)[\kappa(\rho_{11}+2\rho_{13}+\rho_{13})\nonumber \\ &&-2\gamma(\rho_{11}-2\rho_{22}+\rho_{33})]), \nonumber \\
\mathcal{L}_\text{F}\rho_{23} &=& -(2(\gamma+\mu)+\kappa)\rho_{23}-\text {i}\nu(\rho_{33}-\rho_{22}+\rho_{13})\nonumber \\
&& + \frac{\text i f\nu}{4\Omega^2}\Big((2\gamma+\kappa)^2\rho_{22}-(-2\gamma+\kappa)^2\rho_{33} \nonumber \\ &&-\frac{\text i \nu}{4}(f-4)[\kappa\rho_{12}^*+8\text {i}\gamma\Im(2\rho_{23}-\rho_{12})]\Big).
\end{eqnarray}


\bibliographystyle{apsrev4-1}
\bibliography{driving_assisted_quantum_transport}

\providecommand{\noopsort}[1]{}\providecommand{\singleletter}[1]{#1}%
\begin{thebibliography}{35}%
\makeatletter
\providecommand \@ifxundefined [1]{%
 \@ifx{#1\undefined}
}%
\providecommand \@ifnum [1]{%
 \ifnum #1\expandafter \@firstoftwo
 \else \expandafter \@secondoftwo
 \fi
}%
\providecommand \@ifx [1]{%
 \ifx #1\expandafter \@firstoftwo
 \else \expandafter \@secondoftwo
 \fi
}%
\providecommand \natexlab [1]{#1}%
\providecommand \enquote  [1]{``#1''}%
\providecommand \bibnamefont  [1]{#1}%
\providecommand \bibfnamefont [1]{#1}%
\providecommand \citenamefont [1]{#1}%
\providecommand \href@noop [0]{\@secondoftwo}%
\providecommand \href [0]{\begingroup \@sanitize@url \@href}%
\providecommand \@href[1]{\@@startlink{#1}\@@href}%
\providecommand \@@href[1]{\endgroup#1\@@endlink}%
\providecommand \@sanitize@url [0]{\catcode `\\12\catcode `\$12\catcode
  `\&12\catcode `\#12\catcode `\^12\catcode `\_12\catcode `\%12\relax}%
\providecommand \@@startlink[1]{}%
\providecommand \@@endlink[0]{}%
\providecommand \url  [0]{\begingroup\@sanitize@url \@url }%
\providecommand \@url [1]{\endgroup\@href {#1}{\urlprefix }}%
\providecommand \urlprefix  [0]{URL }%
\providecommand \Eprint [0]{\href }%
\providecommand \doibase [0]{http://dx.doi.org/}%
\providecommand \selectlanguage [0]{\@gobble}%
\providecommand \bibinfo  [0]{\@secondoftwo}%
\providecommand \bibfield  [0]{\@secondoftwo}%
\providecommand \translation [1]{[#1]}%
\providecommand \BibitemOpen [0]{}%
\providecommand \bibitemStop [0]{}%
\providecommand \bibitemNoStop [0]{.\EOS\space}%
\providecommand \EOS [0]{\spacefactor3000\relax}%
\providecommand \BibitemShut  [1]{\csname bibitem#1\endcsname}%
\let\auto@bib@innerbib\@empty
\bibitem [{\citenamefont {Liu}\ \emph {et~al.}(2013)\citenamefont {Liu},
  \citenamefont {Levchenko},\ and\ \citenamefont {Baranger}}]{liu2013floquet}%
  \BibitemOpen
  \bibfield  {author} {\bibinfo {author} {\bibfnamefont {D.~E.}\ \bibnamefont
  {Liu}}, \bibinfo {author} {\bibfnamefont {A.}~\bibnamefont {Levchenko}}, \
  and\ \bibinfo {author} {\bibfnamefont {H.~U.}\ \bibnamefont {Baranger}},\
  }\href@noop {} {\bibfield  {journal} {\bibinfo  {journal} {Phys.\ Rev.
  Lett.}\ }\textbf {\bibinfo {volume} {111}},\ \bibinfo {pages} {047002}
  (\bibinfo {year} {2013})}\BibitemShut {NoStop}%
\bibitem [{\citenamefont {Haddadfarshi}\ \emph {et~al.}(2015)\citenamefont
  {Haddadfarshi}, \citenamefont {Cui},\ and\ \citenamefont
  {Mintert}}]{haddadfarshi2015completely}%
  \BibitemOpen
  \bibfield  {author} {\bibinfo {author} {\bibfnamefont {F.}~\bibnamefont
  {Haddadfarshi}}, \bibinfo {author} {\bibfnamefont {J.}~\bibnamefont {Cui}}, \
  and\ \bibinfo {author} {\bibfnamefont {F.}~\bibnamefont {Mintert}},\
  }\href@noop {} {\bibfield  {journal} {\bibinfo  {journal} {Phys.\ Rev.
  Lett.}\ }\textbf {\bibinfo {volume} {114}},\ \bibinfo {pages} {130402}
  (\bibinfo {year} {2015})}\BibitemShut {NoStop}%
\bibitem [{\citenamefont {Eisert}\ \emph {et~al.}(2015)\citenamefont {Eisert},
  \citenamefont {Friesdorf},\ and\ \citenamefont
  {Gogolin}}]{eisert2015quantum}%
  \BibitemOpen
  \bibfield  {author} {\bibinfo {author} {\bibfnamefont {J.}~\bibnamefont
  {Eisert}}, \bibinfo {author} {\bibfnamefont {M.}~\bibnamefont {Friesdorf}}, \
  and\ \bibinfo {author} {\bibfnamefont {C.}~\bibnamefont {Gogolin}},\
  }\href@noop {} {\bibfield  {journal} {\bibinfo  {journal} {Nat. Phys.}\
  }\textbf {\bibinfo {volume} {11}},\ \bibinfo {pages} {124} (\bibinfo {year}
  {2015})}\BibitemShut {NoStop}%
\bibitem [{\citenamefont {Sieberer}\ \emph {et~al.}(2016)\citenamefont
  {Sieberer}, \citenamefont {Buchhold},\ and\ \citenamefont
  {Diehl}}]{sieberer2016keldysh}%
  \BibitemOpen
  \bibfield  {author} {\bibinfo {author} {\bibfnamefont {L.~M.}\ \bibnamefont
  {Sieberer}}, \bibinfo {author} {\bibfnamefont {M.}~\bibnamefont {Buchhold}},
  \ and\ \bibinfo {author} {\bibfnamefont {S.}~\bibnamefont {Diehl}},\
  }\href@noop {} {\bibfield  {journal} {\bibinfo  {journal} {Rep. Prog. Phys.}\
  }\textbf {\bibinfo {volume} {79}},\ \bibinfo {pages} {096001} (\bibinfo
  {year} {2016})}\BibitemShut {NoStop}%
\bibitem [{\citenamefont {Restrepo}\ \emph {et~al.}(2016)\citenamefont
  {Restrepo}, \citenamefont {Cerrillo}, \citenamefont {Bastidas}, \citenamefont
  {Angelakis},\ and\ \citenamefont {Brandes}}]{restrepo2016driven}%
  \BibitemOpen
  \bibfield  {author} {\bibinfo {author} {\bibfnamefont {S.}~\bibnamefont
  {Restrepo}}, \bibinfo {author} {\bibfnamefont {J.}~\bibnamefont {Cerrillo}},
  \bibinfo {author} {\bibfnamefont {V.~M.}\ \bibnamefont {Bastidas}}, \bibinfo
  {author} {\bibfnamefont {D.~G.}\ \bibnamefont {Angelakis}}, \ and\ \bibinfo
  {author} {\bibfnamefont {T.}~\bibnamefont {Brandes}},\ }\href@noop {}
  {\bibfield  {journal} {\bibinfo  {journal} {Phys.\ Rev. Lett.}\ }\textbf
  {\bibinfo {volume} {117}},\ \bibinfo {pages} {250401} (\bibinfo {year}
  {2016})}\BibitemShut {NoStop}%
\bibitem [{\citenamefont {Shu}\ \emph {et~al.}(2018)\citenamefont {Shu},
  \citenamefont {Liu}, \citenamefont {Cao}, \citenamefont {Yang}, \citenamefont
  {Zhang}, \citenamefont {Plenio}, \citenamefont {Jelezko},\ and\ \citenamefont
  {Cai}}]{shu2018observation}%
  \BibitemOpen
  \bibfield  {author} {\bibinfo {author} {\bibfnamefont {Z.}~\bibnamefont
  {Shu}}, \bibinfo {author} {\bibfnamefont {Y.}~\bibnamefont {Liu}}, \bibinfo
  {author} {\bibfnamefont {Q.}~\bibnamefont {Cao}}, \bibinfo {author}
  {\bibfnamefont {P.}~\bibnamefont {Yang}}, \bibinfo {author} {\bibfnamefont
  {S.}~\bibnamefont {Zhang}}, \bibinfo {author} {\bibfnamefont {M.~B.}\
  \bibnamefont {Plenio}}, \bibinfo {author} {\bibfnamefont {F.}~\bibnamefont
  {Jelezko}}, \ and\ \bibinfo {author} {\bibfnamefont {J.}~\bibnamefont
  {Cai}},\ }\href@noop {} {\bibfield  {journal} {\bibinfo  {journal} {Phys.\
  Rev. Lett.}\ }\textbf {\bibinfo {volume} {121}},\ \bibinfo {pages} {210501}
  (\bibinfo {year} {2018})}\BibitemShut {NoStop}%
\bibitem [{\citenamefont {Dai}\ \emph {et~al.}(2016)\citenamefont {Dai},
  \citenamefont {Shi},\ and\ \citenamefont {Yi}}]{dai2016floquet}%
  \BibitemOpen
  \bibfield  {author} {\bibinfo {author} {\bibfnamefont {C.}~\bibnamefont
  {Dai}}, \bibinfo {author} {\bibfnamefont {Z.}~\bibnamefont {Shi}}, \ and\
  \bibinfo {author} {\bibfnamefont {X.}~\bibnamefont {Yi}},\ }\href@noop {}
  {\bibfield  {journal} {\bibinfo  {journal} {Phys.\ Rev. A}\ }\textbf
  {\bibinfo {volume} {93}},\ \bibinfo {pages} {032121} (\bibinfo {year}
  {2016})}\BibitemShut {NoStop}%
\bibitem [{\citenamefont {Behzadi}\ \emph {et~al.}(2015)\citenamefont
  {Behzadi}, \citenamefont {Ahansaz},\ and\ \citenamefont
  {Kasani}}]{behzadi2015dephasing}%
  \BibitemOpen
  \bibfield  {author} {\bibinfo {author} {\bibfnamefont {N.}~\bibnamefont
  {Behzadi}}, \bibinfo {author} {\bibfnamefont {B.}~\bibnamefont {Ahansaz}}, \
  and\ \bibinfo {author} {\bibfnamefont {H.}~\bibnamefont {Kasani}},\
  }\href@noop {} {\bibfield  {journal} {\bibinfo  {journal} {Phys.\ Rev. E}\
  }\textbf {\bibinfo {volume} {92}},\ \bibinfo {pages} {042103} (\bibinfo
  {year} {2015})}\BibitemShut {NoStop}%
\bibitem [{\citenamefont {Barmettler}\ \emph {et~al.}(2008)\citenamefont
  {Barmettler}, \citenamefont {Rey}, \citenamefont {Demler}, \citenamefont
  {Lukin}, \citenamefont {Bloch},\ and\ \citenamefont
  {Gritsev}}]{barmettler2008quantum}%
  \BibitemOpen
  \bibfield  {author} {\bibinfo {author} {\bibfnamefont {P.}~\bibnamefont
  {Barmettler}}, \bibinfo {author} {\bibfnamefont {A.~M.}\ \bibnamefont {Rey}},
  \bibinfo {author} {\bibfnamefont {E.}~\bibnamefont {Demler}}, \bibinfo
  {author} {\bibfnamefont {M.~D.}\ \bibnamefont {Lukin}}, \bibinfo {author}
  {\bibfnamefont {I.}~\bibnamefont {Bloch}}, \ and\ \bibinfo {author}
  {\bibfnamefont {V.}~\bibnamefont {Gritsev}},\ }\href@noop {} {\bibfield
  {journal} {\bibinfo  {journal} {Phys.\ Rev. A}\ }\textbf {\bibinfo {volume}
  {78}},\ \bibinfo {pages} {012330} (\bibinfo {year} {2008})}\BibitemShut
  {NoStop}%
\bibitem [{\citenamefont {Wall}(2015)}]{wall2015quantum}%
  \BibitemOpen
  \bibfield  {author} {\bibinfo {author} {\bibfnamefont {M.~L.}\ \bibnamefont
  {Wall}},\ }\href@noop {} {\emph {\bibinfo {title} {Quantum many-body physics
  of ultracold molecules in optical lattices: Models and simulation methods}}}\
  (\bibinfo  {publisher} {Springer},\ \bibinfo {year} {2015})\BibitemShut
  {NoStop}%
\bibitem [{\citenamefont {Desbuquois}\ \emph {et~al.}(2017)\citenamefont
  {Desbuquois}, \citenamefont {Messer}, \citenamefont {G{\"o}rg}, \citenamefont
  {Sandholzer}, \citenamefont {Jotzu},\ and\ \citenamefont
  {Esslinger}}]{desbuquois2017controlling}%
  \BibitemOpen
  \bibfield  {author} {\bibinfo {author} {\bibfnamefont {R.}~\bibnamefont
  {Desbuquois}}, \bibinfo {author} {\bibfnamefont {M.}~\bibnamefont {Messer}},
  \bibinfo {author} {\bibfnamefont {F.}~\bibnamefont {G{\"o}rg}}, \bibinfo
  {author} {\bibfnamefont {K.}~\bibnamefont {Sandholzer}}, \bibinfo {author}
  {\bibfnamefont {G.}~\bibnamefont {Jotzu}}, \ and\ \bibinfo {author}
  {\bibfnamefont {T.}~\bibnamefont {Esslinger}},\ }\href@noop {} {\bibfield
  {journal} {\bibinfo  {journal} {Phys.\ Rev. A}\ }\textbf {\bibinfo {volume}
  {96}},\ \bibinfo {pages} {053602} (\bibinfo {year} {2017})}\BibitemShut
  {NoStop}%
\bibitem [{\citenamefont {Kyriienko}\ and\ \citenamefont
  {S{\o}rensen}(2018)}]{kyriienko2018floquet}%
  \BibitemOpen
  \bibfield  {author} {\bibinfo {author} {\bibfnamefont {O.}~\bibnamefont
  {Kyriienko}}\ and\ \bibinfo {author} {\bibfnamefont {A.~S.}\ \bibnamefont
  {S{\o}rensen}},\ }\href@noop {} {\bibfield  {journal} {\bibinfo  {journal}
  {Phys.\ Rev. Appl.}\ }\textbf {\bibinfo {volume} {9}},\ \bibinfo {pages}
  {064029} (\bibinfo {year} {2018})}\BibitemShut {NoStop}%
\bibitem [{\citenamefont {Plenio}\ and\ \citenamefont
  {Huelga}(2008)}]{plenio2008dephasing}%
  \BibitemOpen
  \bibfield  {author} {\bibinfo {author} {\bibfnamefont {M.~B.}\ \bibnamefont
  {Plenio}}\ and\ \bibinfo {author} {\bibfnamefont {S.~F.}\ \bibnamefont
  {Huelga}},\ }\href@noop {} {\bibfield  {journal} {\bibinfo  {journal} {New J.
  Phys.}\ }\textbf {\bibinfo {volume} {10}},\ \bibinfo {pages} {113019}
  (\bibinfo {year} {2008})}\BibitemShut {NoStop}%
\bibitem [{\citenamefont {Rebentrost}\ \emph {et~al.}(2009)\citenamefont
  {Rebentrost}, \citenamefont {Mohseni}, \citenamefont {Kassal}, \citenamefont
  {Lloyd},\ and\ \citenamefont {Aspuru-Guzik}}]{rebentrost2009environment}%
  \BibitemOpen
  \bibfield  {author} {\bibinfo {author} {\bibfnamefont {P.}~\bibnamefont
  {Rebentrost}}, \bibinfo {author} {\bibfnamefont {M.}~\bibnamefont {Mohseni}},
  \bibinfo {author} {\bibfnamefont {I.}~\bibnamefont {Kassal}}, \bibinfo
  {author} {\bibfnamefont {S.}~\bibnamefont {Lloyd}}, \ and\ \bibinfo {author}
  {\bibfnamefont {A.}~\bibnamefont {Aspuru-Guzik}},\ }\href@noop {} {\bibfield
  {journal} {\bibinfo  {journal} {New J. Phys.}\ }\textbf {\bibinfo {volume}
  {11}},\ \bibinfo {pages} {033003} (\bibinfo {year} {2009})}\BibitemShut
  {NoStop}%
\bibitem [{\citenamefont {Chin}\ \emph {et~al.}(2012)\citenamefont {Chin},
  \citenamefont {Huelga},\ and\ \citenamefont {Plenio}}]{chin2012coherence}%
  \BibitemOpen
  \bibfield  {author} {\bibinfo {author} {\bibfnamefont {A.}~\bibnamefont
  {Chin}}, \bibinfo {author} {\bibfnamefont {S.}~\bibnamefont {Huelga}}, \ and\
  \bibinfo {author} {\bibfnamefont {M.~B.}\ \bibnamefont {Plenio}},\
  }\href@noop {} {\bibfield  {journal} {\bibinfo  {journal} {Philos. Trans.
  Royal Soc. A}\ }\textbf {\bibinfo {volume} {370}},\ \bibinfo {pages} {3638}
  (\bibinfo {year} {2012})}\BibitemShut {NoStop}%
\bibitem [{\citenamefont {Kassal}\ and\ \citenamefont
  {Aspuru-Guzik}(2012)}]{kassal2012environment}%
  \BibitemOpen
  \bibfield  {author} {\bibinfo {author} {\bibfnamefont {I.}~\bibnamefont
  {Kassal}}\ and\ \bibinfo {author} {\bibfnamefont {A.}~\bibnamefont
  {Aspuru-Guzik}},\ }\href@noop {} {\bibfield  {journal} {\bibinfo  {journal}
  {New J. Phys.}\ }\textbf {\bibinfo {volume} {14}},\ \bibinfo {pages} {053041}
  (\bibinfo {year} {2012})}\BibitemShut {NoStop}%
\bibitem [{\citenamefont {Jang}\ and\ \citenamefont
  {Mennucci}(2018)}]{jang2018delocalized}%
  \BibitemOpen
  \bibfield  {author} {\bibinfo {author} {\bibfnamefont {S.~J.}\ \bibnamefont
  {Jang}}\ and\ \bibinfo {author} {\bibfnamefont {B.}~\bibnamefont
  {Mennucci}},\ }\href@noop {} {\bibfield  {journal} {\bibinfo  {journal} {Rev.
  Mod. Phys.}\ }\textbf {\bibinfo {volume} {90}},\ \bibinfo {pages} {035003}
  (\bibinfo {year} {2018})}\BibitemShut {NoStop}%
\bibitem [{\citenamefont {Mohseni}\ \emph {et~al.}(2008)\citenamefont
  {Mohseni}, \citenamefont {Rebentrost}, \citenamefont {Lloyd},\ and\
  \citenamefont {Aspuru-Guzik}}]{mohseni2008environment}%
  \BibitemOpen
  \bibfield  {author} {\bibinfo {author} {\bibfnamefont {M.}~\bibnamefont
  {Mohseni}}, \bibinfo {author} {\bibfnamefont {P.}~\bibnamefont {Rebentrost}},
  \bibinfo {author} {\bibfnamefont {S.}~\bibnamefont {Lloyd}}, \ and\ \bibinfo
  {author} {\bibfnamefont {A.}~\bibnamefont {Aspuru-Guzik}},\ }\href@noop {}
  {\bibfield  {journal} {\bibinfo  {journal} {J. Chem. Phys.}\ }\textbf
  {\bibinfo {volume} {129}},\ \bibinfo {pages} {11B603} (\bibinfo {year}
  {2008})}\BibitemShut {NoStop}%
\bibitem [{\citenamefont {Li}\ \emph {et~al.}(2015)\citenamefont {Li},
  \citenamefont {Caruso}, \citenamefont {Gauger},\ and\ \citenamefont
  {Benjamin}}]{li2015momentum}%
  \BibitemOpen
  \bibfield  {author} {\bibinfo {author} {\bibfnamefont {Y.}~\bibnamefont
  {Li}}, \bibinfo {author} {\bibfnamefont {F.}~\bibnamefont {Caruso}}, \bibinfo
  {author} {\bibfnamefont {E.}~\bibnamefont {Gauger}}, \ and\ \bibinfo {author}
  {\bibfnamefont {S.~C.}\ \bibnamefont {Benjamin}},\ }\href@noop {} {\bibfield
  {journal} {\bibinfo  {journal} {New Journal of Physics}\ }\textbf {\bibinfo
  {volume} {17}},\ \bibinfo {pages} {013057} (\bibinfo {year}
  {2015})}\BibitemShut {NoStop}%
\bibitem [{\citenamefont {Engel}\ \emph {et~al.}(2007)\citenamefont {Engel},
  \citenamefont {Calhoun}, \citenamefont {Read}, \citenamefont {Ahn},
  \citenamefont {Man{\v{c}}al}, \citenamefont {Cheng}, \citenamefont
  {Blankenship},\ and\ \citenamefont {Fleming}}]{engel2007evidence}%
  \BibitemOpen
  \bibfield  {author} {\bibinfo {author} {\bibfnamefont {G.~S.}\ \bibnamefont
  {Engel}}, \bibinfo {author} {\bibfnamefont {T.~R.}\ \bibnamefont {Calhoun}},
  \bibinfo {author} {\bibfnamefont {E.~L.}\ \bibnamefont {Read}}, \bibinfo
  {author} {\bibfnamefont {T.-K.}\ \bibnamefont {Ahn}}, \bibinfo {author}
  {\bibfnamefont {T.}~\bibnamefont {Man{\v{c}}al}}, \bibinfo {author}
  {\bibfnamefont {Y.-C.}\ \bibnamefont {Cheng}}, \bibinfo {author}
  {\bibfnamefont {R.~E.}\ \bibnamefont {Blankenship}}, \ and\ \bibinfo {author}
  {\bibfnamefont {G.~R.}\ \bibnamefont {Fleming}},\ }\href@noop {} {\bibfield
  {journal} {\bibinfo  {journal} {Nature}\ }\textbf {\bibinfo {volume} {446}},\
  \bibinfo {pages} {782} (\bibinfo {year} {2007})}\BibitemShut {NoStop}%
\bibitem [{\citenamefont {Panitchayangkoon}\ \emph {et~al.}(2010)\citenamefont
  {Panitchayangkoon}, \citenamefont {Hayes}, \citenamefont {Fransted},
  \citenamefont {Caram}, \citenamefont {Harel}, \citenamefont {Wen},
  \citenamefont {Blankenship},\ and\ \citenamefont
  {Engel}}]{panitchayangkoon2010long}%
  \BibitemOpen
  \bibfield  {author} {\bibinfo {author} {\bibfnamefont {G.}~\bibnamefont
  {Panitchayangkoon}}, \bibinfo {author} {\bibfnamefont {D.}~\bibnamefont
  {Hayes}}, \bibinfo {author} {\bibfnamefont {K.~A.}\ \bibnamefont {Fransted}},
  \bibinfo {author} {\bibfnamefont {J.~R.}\ \bibnamefont {Caram}}, \bibinfo
  {author} {\bibfnamefont {E.}~\bibnamefont {Harel}}, \bibinfo {author}
  {\bibfnamefont {J.}~\bibnamefont {Wen}}, \bibinfo {author} {\bibfnamefont
  {R.~E.}\ \bibnamefont {Blankenship}}, \ and\ \bibinfo {author} {\bibfnamefont
  {G.~S.}\ \bibnamefont {Engel}},\ }\href@noop {} {\bibfield  {journal}
  {\bibinfo  {journal} {Proc. Natl. Acad. Sci.}\ }\textbf {\bibinfo {volume}
  {107}},\ \bibinfo {pages} {12766} (\bibinfo {year} {2010})}\BibitemShut
  {NoStop}%
\bibitem [{\citenamefont {Duan}\ \emph {et~al.}(2017)\citenamefont {Duan},
  \citenamefont {Prokhorenko}, \citenamefont {Cogdell}, \citenamefont {Ashraf},
  \citenamefont {Stevens}, \citenamefont {Thorwart},\ and\ \citenamefont
  {Miller}}]{duan2017nature}%
  \BibitemOpen
  \bibfield  {author} {\bibinfo {author} {\bibfnamefont {H.-G.}\ \bibnamefont
  {Duan}}, \bibinfo {author} {\bibfnamefont {V.~I.}\ \bibnamefont
  {Prokhorenko}}, \bibinfo {author} {\bibfnamefont {R.~J.}\ \bibnamefont
  {Cogdell}}, \bibinfo {author} {\bibfnamefont {K.}~\bibnamefont {Ashraf}},
  \bibinfo {author} {\bibfnamefont {A.~L.}\ \bibnamefont {Stevens}}, \bibinfo
  {author} {\bibfnamefont {M.}~\bibnamefont {Thorwart}}, \ and\ \bibinfo
  {author} {\bibfnamefont {R.~D.}\ \bibnamefont {Miller}},\ }\href@noop {}
  {\bibfield  {journal} {\bibinfo  {journal} {Proc. Natl. Acad. Sci.}\ }\textbf
  {\bibinfo {volume} {114}},\ \bibinfo {pages} {8493} (\bibinfo {year}
  {2017})}\BibitemShut {NoStop}%
\bibitem [{\citenamefont {Sowa}\ \emph {et~al.}(2017)\citenamefont {Sowa},
  \citenamefont {Mol}, \citenamefont {Briggs},\ and\ \citenamefont
  {Gauger}}]{sowa2017environment}%
  \BibitemOpen
  \bibfield  {author} {\bibinfo {author} {\bibfnamefont {J.~K.}\ \bibnamefont
  {Sowa}}, \bibinfo {author} {\bibfnamefont {J.~A.}\ \bibnamefont {Mol}},
  \bibinfo {author} {\bibfnamefont {G.~A.~D.}\ \bibnamefont {Briggs}}, \ and\
  \bibinfo {author} {\bibfnamefont {E.~M.}\ \bibnamefont {Gauger}},\
  }\href@noop {} {\bibfield  {journal} {\bibinfo  {journal} {Physical Chemistry
  Chemical Physics}\ }\textbf {\bibinfo {volume} {19}},\ \bibinfo {pages}
  {29534} (\bibinfo {year} {2017})}\BibitemShut {NoStop}%
\bibitem [{\citenamefont {Biggerstaff}\ \emph {et~al.}(2016)\citenamefont
  {Biggerstaff}, \citenamefont {Heilmann}, \citenamefont {Zecevik},
  \citenamefont {Gr{\"a}fe}, \citenamefont {Broome}, \citenamefont {Fedrizzi},
  \citenamefont {Nolte}, \citenamefont {Szameit}, \citenamefont {White},\ and\
  \citenamefont {Kassal}}]{biggerstaff2016enhancing}%
  \BibitemOpen
  \bibfield  {author} {\bibinfo {author} {\bibfnamefont {D.~N.}\ \bibnamefont
  {Biggerstaff}}, \bibinfo {author} {\bibfnamefont {R.}~\bibnamefont
  {Heilmann}}, \bibinfo {author} {\bibfnamefont {A.~A.}\ \bibnamefont
  {Zecevik}}, \bibinfo {author} {\bibfnamefont {M.}~\bibnamefont {Gr{\"a}fe}},
  \bibinfo {author} {\bibfnamefont {M.~A.}\ \bibnamefont {Broome}}, \bibinfo
  {author} {\bibfnamefont {A.}~\bibnamefont {Fedrizzi}}, \bibinfo {author}
  {\bibfnamefont {S.}~\bibnamefont {Nolte}}, \bibinfo {author} {\bibfnamefont
  {A.}~\bibnamefont {Szameit}}, \bibinfo {author} {\bibfnamefont {A.~G.}\
  \bibnamefont {White}}, \ and\ \bibinfo {author} {\bibfnamefont
  {I.}~\bibnamefont {Kassal}},\ }\href@noop {} {\bibfield  {journal} {\bibinfo
  {journal} {Nat. Comm.}\ }\textbf {\bibinfo {volume} {7}},\ \bibinfo {pages}
  {11282} (\bibinfo {year} {2016})}\BibitemShut {NoStop}%
\bibitem [{\citenamefont {Trautmann}\ and\ \citenamefont
  {Hauke}(2018)}]{trautmann2018trapped}%
  \BibitemOpen
  \bibfield  {author} {\bibinfo {author} {\bibfnamefont {N.}~\bibnamefont
  {Trautmann}}\ and\ \bibinfo {author} {\bibfnamefont {P.}~\bibnamefont
  {Hauke}},\ }\href@noop {} {\bibfield  {journal} {\bibinfo  {journal}
  {Physical Review A}\ }\textbf {\bibinfo {volume} {97}},\ \bibinfo {pages}
  {023606} (\bibinfo {year} {2018})}\BibitemShut {NoStop}%
\bibitem [{\citenamefont {Schempp}\ \emph {et~al.}(2015)\citenamefont
  {Schempp}, \citenamefont {G{\"u}nter}, \citenamefont {W{\"u}ster},
  \citenamefont {Weidem{\"u}ller},\ and\ \citenamefont
  {Whitlock}}]{schempp2015correlated}%
  \BibitemOpen
  \bibfield  {author} {\bibinfo {author} {\bibfnamefont {H.}~\bibnamefont
  {Schempp}}, \bibinfo {author} {\bibfnamefont {G.}~\bibnamefont {G{\"u}nter}},
  \bibinfo {author} {\bibfnamefont {S.}~\bibnamefont {W{\"u}ster}}, \bibinfo
  {author} {\bibfnamefont {M.}~\bibnamefont {Weidem{\"u}ller}}, \ and\ \bibinfo
  {author} {\bibfnamefont {S.}~\bibnamefont {Whitlock}},\ }\href@noop {}
  {\bibfield  {journal} {\bibinfo  {journal} {Physical review letters}\
  }\textbf {\bibinfo {volume} {115}},\ \bibinfo {pages} {093002} (\bibinfo
  {year} {2015})}\BibitemShut {NoStop}%
\bibitem [{\citenamefont {Maier}\ \emph {et~al.}(2019)\citenamefont {Maier},
  \citenamefont {Brydges}, \citenamefont {Jurcevic}, \citenamefont {Trautmann},
  \citenamefont {Hempel}, \citenamefont {Lanyon}, \citenamefont {Hauke},
  \citenamefont {Blatt},\ and\ \citenamefont {Roos}}]{maier2019environment}%
  \BibitemOpen
  \bibfield  {author} {\bibinfo {author} {\bibfnamefont {C.}~\bibnamefont
  {Maier}}, \bibinfo {author} {\bibfnamefont {T.}~\bibnamefont {Brydges}},
  \bibinfo {author} {\bibfnamefont {P.}~\bibnamefont {Jurcevic}}, \bibinfo
  {author} {\bibfnamefont {N.}~\bibnamefont {Trautmann}}, \bibinfo {author}
  {\bibfnamefont {C.}~\bibnamefont {Hempel}}, \bibinfo {author} {\bibfnamefont
  {B.~P.}\ \bibnamefont {Lanyon}}, \bibinfo {author} {\bibfnamefont
  {P.}~\bibnamefont {Hauke}}, \bibinfo {author} {\bibfnamefont
  {R.}~\bibnamefont {Blatt}}, \ and\ \bibinfo {author} {\bibfnamefont {C.~F.}\
  \bibnamefont {Roos}},\ }\href@noop {} {\bibfield  {journal} {\bibinfo
  {journal} {Phys.\ Rev. Lett.}\ }\textbf {\bibinfo {volume} {122}},\ \bibinfo
  {pages} {050501} (\bibinfo {year} {2019})}\BibitemShut {NoStop}%
\bibitem [{\citenamefont {Lindblad}(1976)}]{lindblad1976generators}%
  \BibitemOpen
  \bibfield  {author} {\bibinfo {author} {\bibfnamefont {G.}~\bibnamefont
  {Lindblad}},\ }\href@noop {} {\bibfield  {journal} {\bibinfo  {journal}
  {Comm. Math. Phys.}\ }\textbf {\bibinfo {volume} {48}},\ \bibinfo {pages}
  {119} (\bibinfo {year} {1976})}\BibitemShut {NoStop}%
\bibitem [{\citenamefont {Gorini}\ \emph {et~al.}(1976)\citenamefont {Gorini},
  \citenamefont {Kossakowski},\ and\ \citenamefont
  {Sudarshan}}]{gorini1976completely}%
  \BibitemOpen
  \bibfield  {author} {\bibinfo {author} {\bibfnamefont {V.}~\bibnamefont
  {Gorini}}, \bibinfo {author} {\bibfnamefont {A.}~\bibnamefont {Kossakowski}},
  \ and\ \bibinfo {author} {\bibfnamefont {E.~C.~G.}\ \bibnamefont
  {Sudarshan}},\ }\href@noop {} {\bibfield  {journal} {\bibinfo  {journal} {J.
  Math. Phys.}\ }\textbf {\bibinfo {volume} {17}},\ \bibinfo {pages} {821}
  (\bibinfo {year} {1976})}\BibitemShut {NoStop}%
\bibitem [{\citenamefont {Blanes}\ \emph {et~al.}(2009)\citenamefont {Blanes},
  \citenamefont {Casas}, \citenamefont {Oteo},\ and\ \citenamefont
  {Ros}}]{blanes2009magnus}%
  \BibitemOpen
  \bibfield  {author} {\bibinfo {author} {\bibfnamefont {S.}~\bibnamefont
  {Blanes}}, \bibinfo {author} {\bibfnamefont {F.}~\bibnamefont {Casas}},
  \bibinfo {author} {\bibfnamefont {J.}~\bibnamefont {Oteo}}, \ and\ \bibinfo
  {author} {\bibfnamefont {J.}~\bibnamefont {Ros}},\ }\href@noop {} {\bibfield
  {journal} {\bibinfo  {journal} {Phys. Rep.}\ }\textbf {\bibinfo {volume}
  {470}},\ \bibinfo {pages} {151} (\bibinfo {year} {2009})}\BibitemShut
  {NoStop}%
\bibitem [{\citenamefont {Johansson}\ \emph {et~al.}(2013)\citenamefont
  {Johansson}, \citenamefont {Nation},\ and\ \citenamefont
  {Nori}}]{johansson2013qutip}%
  \BibitemOpen
  \bibfield  {author} {\bibinfo {author} {\bibfnamefont {J.~R.}\ \bibnamefont
  {Johansson}}, \bibinfo {author} {\bibfnamefont {P.~D.}\ \bibnamefont
  {Nation}}, \ and\ \bibinfo {author} {\bibfnamefont {F.}~\bibnamefont
  {Nori}},\ }\href@noop {} {\bibfield  {journal} {\bibinfo  {journal} {Comput.
  Phys. Commun.}\ }\textbf {\bibinfo {volume} {184}},\ \bibinfo {pages} {1234}
  (\bibinfo {year} {2013})}\BibitemShut {NoStop}%
\bibitem [{\citenamefont {Anderson}(1958)}]{anderson1958absence}%
  \BibitemOpen
  \bibfield  {author} {\bibinfo {author} {\bibfnamefont {P.~W.}\ \bibnamefont
  {Anderson}},\ }\href@noop {} {\bibfield  {journal} {\bibinfo  {journal}
  {Phys.\ Rev.}\ }\textbf {\bibinfo {volume} {109}},\ \bibinfo {pages} {1492}
  (\bibinfo {year} {1958})}\BibitemShut {NoStop}%
\bibitem [{\citenamefont {Hofer}\ \emph {et~al.}(2017)\citenamefont {Hofer},
  \citenamefont {Perarnau-Llobet}, \citenamefont {Miranda}, \citenamefont
  {Haack}, \citenamefont {Silva}, \citenamefont {Brask},\ and\ \citenamefont
  {Brunner}}]{hofer2017markovian}%
  \BibitemOpen
  \bibfield  {author} {\bibinfo {author} {\bibfnamefont {P.~P.}\ \bibnamefont
  {Hofer}}, \bibinfo {author} {\bibfnamefont {M.}~\bibnamefont
  {Perarnau-Llobet}}, \bibinfo {author} {\bibfnamefont {L.~D.~M.}\ \bibnamefont
  {Miranda}}, \bibinfo {author} {\bibfnamefont {G.}~\bibnamefont {Haack}},
  \bibinfo {author} {\bibfnamefont {R.}~\bibnamefont {Silva}}, \bibinfo
  {author} {\bibfnamefont {J.~B.}\ \bibnamefont {Brask}}, \ and\ \bibinfo
  {author} {\bibfnamefont {N.}~\bibnamefont {Brunner}},\ }\href@noop {}
  {\bibfield  {journal} {\bibinfo  {journal} {New Journal of Physics}\ }\textbf
  {\bibinfo {volume} {19}},\ \bibinfo {pages} {123037} (\bibinfo {year}
  {2017})}\BibitemShut {NoStop}%
\bibitem [{\citenamefont {Van Der~Walt}\ \emph {et~al.}(2011)\citenamefont {Van
  Der~Walt}, \citenamefont {Colbert},\ and\ \citenamefont
  {Varoquaux}}]{van2011numpy}%
  \BibitemOpen
  \bibfield  {author} {\bibinfo {author} {\bibfnamefont {S.}~\bibnamefont {Van
  Der~Walt}}, \bibinfo {author} {\bibfnamefont {S.~C.}\ \bibnamefont
  {Colbert}}, \ and\ \bibinfo {author} {\bibfnamefont {G.}~\bibnamefont
  {Varoquaux}},\ }\href@noop {} {\bibfield  {journal} {\bibinfo  {journal}
  {Comput. Sci. Eng.}\ }\textbf {\bibinfo {volume} {13}},\ \bibinfo {pages}
  {22} (\bibinfo {year} {2011})}\BibitemShut {NoStop}%
\bibitem [{\citenamefont {Hunter}(2007)}]{hunter2007matplotlib}%
  \BibitemOpen
  \bibfield  {author} {\bibinfo {author} {\bibfnamefont {J.~D.}\ \bibnamefont
  {Hunter}},\ }\href@noop {} {\bibfield  {journal} {\bibinfo  {journal}
  {Comput. Sci. Eng.}\ }\textbf {\bibinfo {volume} {9}},\ \bibinfo {pages} {90}
  (\bibinfo {year} {2007})}\BibitemShut {NoStop}%
\end{thebibliography}%

\pagebreak
\thispagestyle{plain}
\setcounter{page}{1}
\renewcommand{\theequation}{S\arabic{equation}}

\preprint{APS/123-QED}

\title{Supplemental Material: Driving-assisted open quantum transport in qubit networks}

\maketitle

\onecolumngrid

\centering\large{\textbf{Supplemental Material: Driving-assisted open quantum transport in qubit networks}}

\section*{I. Solution of the Floquet-Magnus-Markov expansion}
Here we write the coefficients $A_1$--$B_6$ for the solution of the analytical on-site transport efficiency in Eq. (22). The corresponding differential equation is Eq. (A1). The solution is obtained via Gaussian elimination. Setting $\nu=1$, the numerator coefficients $A_0$--$A_5$ are
\begin{small}
\begin{eqnarray}
A_0 &=& 128\Omega^{16} \big[2 \gamma + \kappa + 2 \mu\big] \big[\kappa + 2 \gamma (2 + (2 \gamma + \kappa)^2) + 4 \mu +  2 (2 \gamma + \kappa) (6 \gamma + \kappa) \mu + 8 (3 \gamma + \kappa) \mu^2 + 8 \mu^3\big] \nonumber  \big[1 + 
   2 \mu (\gamma + \mu)\big], \nonumber \\
A_1 &=& 64 \Omega^{12} \Big\{-128 \gamma^6 - 192 \gamma^5 \big[\kappa + 2 \mu\big] + 
   4 \gamma \big[\kappa^3 (9 + 4 \mu^2) \nonumber  + \kappa^2 \mu (79 + 24 \mu^2) + 
      8 \kappa (3 + 28 \mu^2 + 6 \mu^4) + 4 \mu (15 \nonumber \\&&+ 55 \mu^2 + 8 \mu^4)\big] - 
   2 \gamma \big[4 \kappa + \kappa^3 + 10 \mu + 11 \kappa^2 \mu \nonumber + 32 \kappa \mu^2 + 32 \mu^3\big] \Omega^2 - 
   8 \mu^2 (1 + 2 \mu^2) (-13 + \Omega^2)\nonumber \\&& - 2 \kappa^3 \mu (-10 + \Omega^2) - 
   16 \gamma^4 \big[-12 + 6 \kappa^2 + 24 \kappa \mu + 16 \mu^2  + \Omega^2\big] - 
   8 \gamma^3 \big[2 (-22 \kappa + \kappa^3 - 53 \mu + 6 \kappa^2 \mu - 16 \mu^3) \nonumber \nonumber \\&& + (3 \kappa +          8 \mu) \Omega^2\big]  + 
   4 \gamma^2 \big[34 + 50 \kappa^2 + 248 \kappa \mu+ 4 (83 + 6 \kappa^2) \mu^2 + 96 \kappa \mu^3 + 
      96 \mu^4 - 3 - (\kappa + 3 \mu) \nonumber (3 \kappa + 8 \mu) \Omega^2\big]  \nonumber \\&&  + 
   \kappa \mu (80 - 7 \Omega^2 + 4 \mu^2 (64 - 5 \Omega^2))- 
   \kappa^2 (-12 + \Omega^2 + 2 \mu^2 (-58 + 5 \Omega^2))\Big\}, \nonumber \\
A_2 &=& 8 \Omega^8 \Big\{768 \gamma^4 - 1024 \gamma^6 + 144 \kappa^2 + 1312 \kappa \mu + 1856 \mu^2  - 
   768 \gamma^4 \big[\kappa + 2 \mu\big]\big[\kappa + 4 \mu\big] - 512 \gamma^5 \big[3 \kappa + 8 \mu\big] \nonumber \\&& + 
   32 \gamma \big[63 \kappa + 8 \kappa^3 + 146 \mu + 42 \kappa^2 \mu + 88 \kappa \mu^2 + 72 \mu^3\big]  - 
   128 \gamma^2 \big[-12 (2 + \kappa^2) + \kappa (-41 + \kappa^2) \mu + 6 (-7 + \kappa^2) \mu^2 \nonumber \\&& + 
      12 \kappa \mu^3 + 8 \mu^4\big] - 
   128 \gamma^3 \big[\kappa^3 - 30 \mu + 12 \kappa^2 \mu + 32 \mu^3  + \kappa (-19 + 36 \mu^2)\big] - 
   16 \Big[38 \gamma^2 + 32 \gamma^4 + 22 \gamma \kappa + 56 \gamma^3 \kappa \nonumber \\&& + \kappa^2 + 30 \gamma^2 \kappa^2 + 
      5 \gamma \kappa^3 + (51 \gamma + 92 \gamma^3 + 13 \kappa + 96 \gamma^2 \kappa  + 21 \gamma \kappa^2 - 
         \kappa^3) \mu + (16 + (38 \gamma - 7 \kappa) (2 \gamma + \kappa)) \mu^2 \nonumber \\&&+ 
      4 (\gamma - 4 \kappa) \mu^3 - 12 \mu^4\Big] \Omega^2 + \Big[16 \gamma^4 + 2 \kappa^3 \mu + 
      8 \gamma^3 (3 \kappa + 8 \mu) + 5 \kappa \mu (3 + 4 \mu^2) + \kappa^2 (1 + 10 \mu^2) + 
      2 \gamma (\kappa (8 + \kappa^2)\nonumber \\&& + 11 (2 + \kappa^2) \mu + 32 \kappa \mu^2 + 32 \mu^3) + 
      16 (\mu^2 + \mu^4) + 4 \gamma^2 \big[7 + (\kappa + 3 \mu) (3 \kappa + 8 \mu)\big]\Big] \Omega^4\Big\}
, \nonumber \\
A_3 &=& 16 \Omega^4 \Big\{-32 \gamma \Big[-7 \kappa + 7 \mu + 
      \gamma \big[-5 + (2 \gamma + \kappa + 2 \mu)(10 \gamma + \kappa + 10 \mu)\big]\Big] - 
   4 \Big[48 \gamma^4 + 88 \gamma^3 (\kappa + 2 \mu) \nonumber \\&& - (\kappa + 3 \mu) (3 \kappa + 16 \mu) + 
      4 \gamma^2 \big[17 + 2 (\kappa + 2 \mu) (6 \kappa + 13 \mu)\big] + 
      4 \gamma \big[2 \kappa^3 - 2 \mu + 13 \kappa^2 \mu + 20 \mu^3 \nonumber  + 
         4 \kappa (2 + 7 \mu^2)\big]\Big] \Omega^2 \nonumber \\&&+ \Big[-\kappa^2 + 
      \kappa (-2 + \kappa^2) \mu + (-11 + 7 \kappa^2) \mu^2 + 16 \kappa \mu^3 + 12 \mu^4 + 
      4 \gamma^3 \big[2 \kappa + 7 \mu\big] + \gamma^2 \big[62 + 6 \kappa^2 + 40 \kappa \mu + 68 \mu^2\big]\nonumber \\&& + 
      \gamma \big[\kappa^3 + 41 \mu + 13 \kappa^2 \mu + 52 \mu^3 + 6 \kappa (5 + 8 \mu^2)\big]\Big] \Omega^4 - \Big[\gamma +
       \mu\Big] \Big[2 \gamma + \kappa + \mu\Big] \Omega^6\Big\}
, \nonumber \\[.1 cm]
A_4 &=& \Omega^2 \Big\{-128 \gamma \big[34 \gamma + 14 \kappa + 35 \mu\big] - 
   8 \Big[16 \gamma^4 + 8 \gamma^3 (\kappa + 4 \mu) - (\kappa + 3 \mu) (\kappa + 10 \mu) - 
      2 \gamma (26 \kappa + 61 \mu) \nonumber \\&&+ 4 \gamma^2 \big[-27 + 2 \mu (\kappa + 2 \mu)\big]\Big] \Omega^2 - 
   16 \gamma \Big[6 \gamma + 3 \kappa + 5 \mu\Big] \Omega^4 + \Big[\gamma + \mu\Big] \Big[2 \gamma + \kappa + \mu\Big] \Omega^6\Big\}
, \nonumber \\[.1 cm]
A_5 &=& -32 \gamma^2 - 
   16 \gamma \Big[4 \gamma + 2 \kappa + 5 \mu\Big] \Omega^2  + \Big[\gamma + \mu\Big] \Big[2 \gamma + \kappa + 3 \mu\Big]\Omega^4,      
\end{eqnarray}
\end{small}
and the denominator coefficients $B_0$--$B_6$ are
\begin{small}
\begin{eqnarray}
B_0 &=& 128 \Omega^{16} \Big\{\kappa + 2 \gamma (2 + (2 \gamma + \kappa)^2) + 4 \mu + 2 (2 \gamma + \kappa) (6 \gamma + \kappa) \mu + 
   8 (3 \gamma + \kappa) \mu^2 + 8 \mu^3 \Big\}  \Big\{8 \gamma^3 \mu^2 (\kappa + \mu) \nonumber \qquad\qquad\qquad\;\;\\&&+ 
   4 \gamma^2 \mu  \big[3 \kappa + 2 (2 + \kappa^2) \mu + 8 \kappa \mu^2 + 6 \mu^3 \big] +  \big[\kappa + 2 \mu + 
      \kappa \mu^2 + \mu^3 \big]  \big[\kappa + 2 (2 + \kappa^2) \mu + 8 \kappa \mu^2 + 8 \mu^3 \big]\nonumber \\&& + 
   2 \gamma  \big[\kappa + (3 + 5 \kappa^2) \mu + \kappa (20 + \kappa^2) \mu^2 + 9 (2 + \kappa^2) \mu^3 + 
      20 \kappa \mu^4 + 12 \mu^5 \big] \Big\}
, \nonumber  \end{eqnarray}\begin{eqnarray}
B_1 &=& 64 \Omega^{12} \Big\{-512 \gamma^7 \mu  \big[\kappa + \mu \big] - 
   128 \gamma^6  \big[\kappa + 3 \mu + 8 \kappa^2 \mu+ 28 \kappa \mu^2 + 20 \mu^3 \big] - 
   64 \gamma^5  \big[12 \kappa^3 \mu + \kappa^2 (3 + 76 \mu^2)\nonumber \\&& + \mu^2 (12 + 80 \mu^2 + \Omega^2) + 
      \kappa \mu (6 + 144 \mu^2 + \Omega^2) \big] -  \big[\kappa + 4 \mu + 2 \kappa^2 \mu + 8 \kappa \mu^2 + 
      8 \mu^3 \big]  \big[12 \mu^2 (1 + \mu^2) (-10 + \Omega^2) \nonumber \\&&+ 2 \kappa^3 \mu (-8 + \Omega^2) + 
      \kappa \mu (9 (-10 + \Omega^2) + 4 \mu^2 (-47 + 5 \Omega^2)) + 
      \kappa^2 (-12 + \Omega^2 + 2 \mu^2 (-44 + 5 \Omega^2)) \big] \nonumber \\&&- 
   16 \gamma^4  \big[16 \kappa^4 \mu + 320 \mu^5 + 2 \kappa^3 (3 + 80 \mu^2) + 
      3 \mu (-12 + \Omega^2) + 22 \mu^3 (-6 + \Omega^2) + 
      4 \kappa^2 \mu (-13 + 132 \mu^2 + 2 \Omega^2)\nonumber \\&& + 
      \kappa (-12 + 704 \mu^4 + \Omega^2 + \mu^2 (-212 + 30 \Omega^2)) \big] - 
   4 \gamma^2  \big[8 \kappa^5 \mu^2 + 2 \kappa^4 \mu (-61 + 36 \mu^2 + 4 \Omega^2) + 
      \mu (-102 \nonumber \\&&+ 128 \mu^6 + 9 \Omega^2 + 16 \mu^4 (-150 + 13 \Omega^2) + 
         8 \mu^2 (-197 + 17 \Omega^2)) + 
      2 \kappa^2 \mu (-300 + 224 \mu^4 + 23 \Omega^2 + 5 \mu^2 (-350 \nonumber \\&& + 29 \Omega^2))+ 
      \kappa^3 (-50 + 256 \mu^4 + 3 \Omega^2 + 2 \mu^2 (-538 + 41 \Omega^2)) \nonumber \\&&+ 
      \kappa (-34 + 384 \mu^6 + 3 \Omega^2 + 12 \mu^2 (-154 + 13 \Omega^2) + 
         8 \mu^4 (-613 + 53 \Omega^2)) \big] - 
   8 \gamma^3  \big[4 \kappa^5 \mu + 320 \mu^6 \nonumber \\&&+ \kappa^4 (2 + 68 \mu^2) + 34 \mu^2 (-12 + \Omega^2) + 
      48 \mu^4 (-21 + 2 \Omega^2) + 4 \kappa^3 \mu (-37 + 88 \mu^2 + 3 \Omega^2) + 
      \kappa^2 (-44 + 800 \mu^4 + 3 \Omega^2\nonumber \\&& + \mu^2 (-916 + 80 \Omega^2)) + 
      2 \kappa \mu (-161 + 416 \mu^4 + 13 \Omega^2  + \mu^2 (-882 + 82 \Omega^2)) \big] - 
   2 \gamma  \Big[2 \kappa^5 \mu (-16 + \Omega^2) \nonumber \\&&+ 
      2 \mu^2 (-230 + 21 \Omega^2 + 16 \mu^4 (-78 + 7 \Omega^2) + 
         4 \mu^2 (-316 + 29 \Omega^2)) + 
      \kappa^4 (-18 + \Omega^2 + \mu^2 (-428 + 34 \Omega^2)) \nonumber \\&&+ 
      \kappa^3 \mu (-346 + 27 \Omega^2 + 4 \mu^2 (-531 + 46 \Omega^2)) + 
      2 \kappa^2 (2 (-12 + \Omega^2) + 4 \mu^4 (-634 + 57 \Omega^2) + 
         \mu^2 (-928 \nonumber  \\&&+ 81 \Omega^2)) + 
      \kappa \mu (-322 + 29 \Omega^2 + 528 \mu^4 (-11 + \Omega^2) + 
         \mu^2 (-3788 + 346 \Omega^2)) \Big]\Big\}
, \nonumber   \\[.1 cm]
B_2 &=& 8 \Omega^8 \Big\{-1024 \gamma^6  \Big[\kappa + 3 \mu \Big] + 4 \kappa^5 \mu^2 \Omega^4 + 
   4 \kappa^4 \mu (64 - 8 \Omega^2\qquad + (1 + 9 \mu^2) \Omega^4) - 
   64 \gamma^5  \Big[24 \kappa^2 + \kappa \mu (280 + 16 \Omega^2 - \Omega^4) \nonumber \\&&+ 
      \mu^2 (336 + 16 \Omega^2 - \Omega^4) \Big] + 
   48 \mu^3 (200 - 40 \Omega^2 + 3 \Omega^4 + 2 \mu^4 \Omega^4 + 
      6 \mu^2 (50 - 10 \Omega^2 + \Omega^4)) + 
   \kappa^3 (144 - 16 \Omega^2 \nonumber \\&&+ \Omega^4 + 136 \mu^4 \Omega^4 + 
      4 \mu^2 (980 - 184 \Omega^2 + 19 \Omega^4)) + 
   \kappa^2 \mu (2304 - 400 \Omega^2 + 29 \Omega^4 + 264 \mu^4 \Omega^4 + 
      4 \mu^2 (4020 \nonumber \\&&- 808 \Omega^2 + 81 \Omega^4)) + 
   4 \kappa \mu^2 (64 \mu^4 \Omega^4 + 12 (200 - 40 \Omega^2 + 3 \Omega^4) + 
      \mu^2 (6560 - 1336 \Omega^2 + 133 \Omega^4)) - 
   8 \gamma^3  \Big[16 \kappa^4 \nonumber \\&&+ 4 \kappa^3 \mu (148 + 88 \Omega^2 - 3 \Omega^4) + 
      2 \kappa \mu (-1576 + 484 \Omega^2 - 21 \Omega^4 + 
         \mu^2 (4064 + 1216 \Omega^2 - 82 \Omega^4)) + 
      \kappa^2 (-304 \nonumber \\&&+ 112 \Omega^2 - 3 \Omega^4 + \mu^2 (3936 + 1632 \Omega^2 - 80 \Omega^4)) + 
      2 \mu^2 (-1808 + 544 \Omega^2 - 25 \Omega^4 - 
         48 \mu^2 (-52 - 12 \Omega^2 \nonumber \\&&+ \Omega^4)) \Big] - 
   16 \gamma^4 \Big[48 \kappa^3 + \mu^3 (2880 + 320 \Omega^2 - 22 \Omega^4) - 
      3 \mu (48 - 32 \Omega^2 + \Omega^4) - 8 \kappa^2 \mu (-130 - 24 \Omega^2 + \Omega^4) \nonumber \\&&- 
      \kappa (48 - 32 \Omega^2 + \Omega^4 + \mu^2 (-3584 - 512 \Omega^2 + 30 \Omega^4))\Big] + 
   4 \gamma^2 \Big[8 \kappa^4 \mu (-8 - 32 \Omega^2 + \Omega^4) + 3 \mu (768 - 152 \Omega^2 + 7 \Omega^4) \nonumber \\&&+ 
      16 \mu^5 (-192 - 112 \Omega^2 + 13 \Omega^4) + 
      4 \mu^3 (4292 - 1004 \Omega^2  + 59 \Omega^4) + 
      \kappa^3 (3 (128 - 40 \Omega^2 + \Omega^4) + 2 \mu^2 (-512 - 864 \Omega^2 \nonumber \\&& + 41 \Omega^4)) +
       \kappa (768 - 152 \Omega^2 + 7 \Omega^4 + 48 \mu^2 (455 - 111 \Omega^2 + 6 \Omega^4) + 
         8 \mu^4 (-768 - 576 \Omega^2 + 53 \Omega^4)) + 
      2 \kappa^2 \mu (3488 \nonumber \\&&- 888 \Omega^2 + 39 \Omega^4 + 
         \mu^2 (-2048 - 2144 \Omega^2 + 145 \Omega^4))\Big] + 
   2 \gamma \Big[2 \kappa^5 \mu \Omega^2 (-32 + \Omega^2) + 
      \kappa^4 (128 - 8 (5 + 72 \mu^2) \Omega^2 \nonumber \\ &&+ (1 + 34 \mu^2) \Omega^4) + 
      \kappa^3 \mu (3808 - 8 (111 + 256 \mu^2) \Omega^2 + (43 + 184 \mu^2) \Omega^4) + 
      2 \kappa^2 (4 \mu^4 \Omega^2 (-448 + 57 \Omega^2) \nonumber \\&&+ 4 (126 - 22 \Omega^2 + \Omega^4) + 
         3 \mu^2 (3696 - 824 \Omega^2 + 51 \Omega^4)) + 
      2 \mu^2 (5216 - 1000 \Omega^2 + 57 \Omega^4 + 16 \mu^4 \Omega^2 (-32 + 7 \Omega^2) \nonumber \\&&+ 
         4 \mu^2 (3528 - 736 \Omega^2 + 55 \Omega^4)) + 
      \kappa \mu (7696 - 1488 \Omega^2 + 81 \Omega^4 + 48 \mu^4 \Omega^2 (-64 + 11 \Omega^2) \nonumber \\&&+ 
         \mu^2 (44896 - 9744 \Omega^2 + 682 \Omega^4))\Big]\Big\}
, \nonumber  \\[.1 cm]
B_3 &=& 16 \Omega^4 \Big\{2 \kappa^4 \mu \Omega^4 - 64 \gamma^5 \mu \big[\kappa + \mu\big] \Omega^4 - 
   \kappa^3 \Omega^2 \big[-12 + (1 - 46 \mu^2) \Omega^2 + 4 \mu^2 \Omega^4\big] - 
   2 \mu^3 \big[-1568 + 384 \Omega^2 \nonumber \\&&- 3 (31 + 30 \mu^2) \Omega^4 + (8 + 9 \mu^2) \Omega^6\big] + 
   \kappa^2 \mu \big[392 - 4 \Omega^2 + (15 + 202 \mu^2) \Omega^4 - 2 (1 + 10 \mu^2) \Omega^6\big] - 
   2 \kappa \mu^2 \big[-1568 \nonumber \\&&+ 
      384 \Omega^2 - (93 + 167 \mu^2) \Omega^4 + (8 + 17 \mu^2) \Omega^6\big] - 
   64 \gamma^4 \Big[\kappa^2 \mu \Omega^4 + 3 \mu (10 + 3 \Omega^2 + \mu^2 \Omega^4) + 
      \kappa (10 + 3 \Omega^2 \nonumber \\&&+ 4 \mu^2 \Omega^4)\Big] - 
   4 \gamma^3 \Big[4 \kappa^3 \mu \Omega^4 + \kappa^2 (96 + 88 \Omega^2 + (-2 + 36 \mu^2) \Omega^4) + 
      4 \mu^2 (400 + 196 \Omega^2 \nonumber \\&&+ 4 (-7 + 3 \mu^2) \Omega^4 + \Omega^6) + 
      \kappa \mu (1248 + 728 \Omega^2+ (-109 + 80 \mu^2) \Omega^4 + 4 \Omega^6)\Big] + 
   \gamma \Big[\kappa^4 \Omega^2 (-32 + \Omega^2) \nonumber
\\
   &&+ \kappa^3 \mu \Omega^2 (-736 + 131 \Omega^2 - 4 \Omega^4) - 
      2 \mu^2 (-3024 + 8 (147 + 160 \mu^2) \Omega^2 - 
         9 (25 + 48 \mu^2) \Omega^4 + (11 + 26 \mu^2) \Omega^6) \nonumber \\&&- 
      \kappa^2 (-224 + 128 (1 + 25 \mu^2) \Omega^2 - 
         2 (15 + 389 \mu^2) \Omega^4  + (1 + 36 \mu^2) \Omega^6) + 
      \kappa \mu (5040 - 8 (265 + 624 \mu^2) \Omega^2 \nonumber \\&&+ 
         2 (191 + 745 \mu^2) \Omega^4 - (17 + 84 \mu^2) \Omega^6)\Big] - 
   2 \gamma^2 \Big[32 \mu^5 \Omega^4+ \kappa^3 (16 + 96 \Omega^2 + (-3 + 8 \mu^2) \Omega^4) + 
      3 \mu (-80 + 136 \Omega^2 \nonumber \\&&- 31 \Omega^4 + \Omega^6) + 
      2 \kappa^2 \mu (184 + 768 \Omega^2 + 5 (-25 + 4 \mu^2) \Omega^4  + 4 \Omega^6) + 
      \mu^3 (2240 + 2560 \Omega^2 - 566 \Omega^4 + 25 \Omega^6) \nonumber \\&&+ 
      \kappa (-80 + 136 \Omega^2 - 31 \Omega^4 + 64 \mu^4 \Omega^4 + \Omega^6 + 
         \mu^2 (2304 + 3840 \Omega^2 - 794 \Omega^4 + 33 \Omega^6))\Big]\Big\}
, \nonumber \end{eqnarray}\begin{eqnarray}
B_4 &=&  \Omega^2 \Big\{-4352 \gamma^2 \big[\kappa + 3 \mu\big] - 1792 \gamma \big[\kappa^2 + 12 \kappa \mu + 13 \mu^2\big] - 
   8 \Big[16 \gamma^4 \big[\kappa + 3 \mu\big] + 
      8 \gamma^3 \big[\kappa^2 + 25 \kappa \mu + 30 \mu^2\big] \nonumber \\&&- (\kappa + 10 \mu) (\kappa^2 + 27 \kappa \mu + 
         30 \mu^2) - 2 \gamma \big[26 \kappa^2 + 521 \kappa \mu + 580 \mu^2\big] + 
      4 \gamma^2 \big[-81 \mu + 4 \kappa^2 \mu + 48 \mu^3 \nonumber \\&&+ 3 \kappa (-9 + 16 \mu^2)\big]\Big] \Omega^2 - 
   16 \Big[24 \gamma^3 \mu (\kappa + \mu) + 2 \mu (\kappa^2 + 15 \kappa \mu + 15 \mu^2) + 
      2 \gamma^2 \big[3 \kappa + 9 \mu + 16 \kappa^2 \mu+ 44 \kappa \mu^2 + 28 \mu^3\big] \nonumber \\&& + 
      \gamma \big[8 \kappa^3 \mu + 4 \mu^2 (19 + 8 \mu^2) + \kappa^2 (3 + 40 \mu^2) + 
         \kappa \mu (69 + 64 \mu^2)\big]\Big] \Omega^4 - \Big[-16 \gamma^3 \mu (\kappa + \mu) - 
      2 \gamma^2 \big[\kappa + 3 \mu + 8 \kappa^2 \mu \nonumber \\&&+ 33 \kappa \mu^2 + 25 \mu^3\big] - 
      2 \mu (\kappa^2+ 2 \kappa (9 + \kappa^2) \mu + 2 (9 + 5 \kappa^2) \mu^2 + 17 \kappa \mu^3 + 
         9 \mu^4)\nonumber \\&& - 
      \gamma \big[\kappa^2 + \kappa (37 + 4 \kappa^2) \mu + 6 (7 + 6 \kappa^2) \mu^2 + 84 \kappa \mu^3 + 
         52 \mu^4\big]\Big] \Omega^6\Big\}
, \nonumber \\
B_5 &=& 32 \gamma^2 \big[\kappa + 3 \mu\big] + 
   32 \gamma \Big[\kappa^2 + 19 \kappa \mu + 20 \mu^2 + 
      2 \gamma (\kappa + 3 \mu)\Big] \Omega^2 + \Big[8 \gamma^3 \mu (\kappa + \mu) - 
      2 \mu (\kappa^2 + 15 \kappa \mu \nonumber \\&&+ 15 \mu^2) - \gamma \big[\kappa^2 + 103 \kappa \mu + 108 \mu^2\big] + 
      \gamma^2 (-6 \mu + 8 \mu^3+ \kappa (-2 + 8 \mu^2))\Big] \Omega^4 + 
   3 \mu \Big[\gamma + \mu\Big] \Big[\kappa + \mu\Big] \Omega^6
, \nonumber \\
B_6 &=& \frac{\Omega^2}{8} \mu \Big[\kappa + \mu\Big]  \Big[-32 \gamma + (\gamma + \mu) \Omega^2\Big].
\end{eqnarray}
The solution converges to $\eta_0$ in Eq. (23) in absence of driving (setting $\Delta=0$ or taking limit $\Omega\rightarrow\infty$).
\end{small}



\end{document}